\documentclass[useAMS,referee,usenatbib,noendfloat]{biom}

\usepackage{amsmath,amssymb,amsfonts,amsbsy}
\usepackage{graphicx}
\usepackage{bbm}
\usepackage{mathtools}
\usepackage{float}
\usepackage{xcolor}
\usepackage{makecell}
\usepackage{hyperref}

\makeatletter
\renewcommand{\newauthor}{%
  \ifnum\c@emailcnt=0\else \printemails\\\setcounter{emailcnt}{0}\fi
  \ifnum\c@emailxcnt=0\else \printemailxs\\\setcounter{emailxcnt}{0}\fi
  \end{tabular}%
  \vskip 0pt 
  \par
  \begin{tabular}[t]{@{}c@{}}\let\\=\author@nextline%
}
\makeatother

\title[
Vaccine sieve analysis with deep sequencing
]{
Vaccine sieve analysis of deep sequencing data using competing risks Cox regression with failure type subject to misclassification
}

\author{
James Peng$^{1}$\email{james.s.peng@berkeley.edu},
Michal Juraska$^{2}$,
Pamela A. Shaw$^{3}$, and
Peter B. Gilbert$^{1,2}$ \\
$^{1}$Department of Biostatistics, University of Washington, Seattle, WA, USA \\
$^{2}$Vaccine and Infectious Disease Division, Fred Hutchinson Cancer Center, Seattle, WA, USA \\
$^{3}$Kaiser Permanente Washington Health Research Institute, Seattle, WA, USA
}

\begin{document}

\label{firstpage}

\begin{abstract}
Understanding how vaccines perform against different pathogen genotypes is crucial for developing effective prevention strategies, particularly for highly genetically diverse pathogens like HIV. Sieve analysis is a statistical framework used to determine whether a vaccine selectively prevents infection by certain genotypes while allowing breakthrough of other genotypes that evade immune responses. Traditionally, these analyses are conducted with a single sequence available per individual acquiring the pathogen. However, modern sequencing technology can provide detailed characterization of intra-individual viral diversity by capturing up to hundreds of pathogen sequences per person. In this work, we introduce methodology that extends sieve analysis to account for intra-individual viral diversity. Our approach estimates vaccine efficacy against viral populations with at least a small threshold of vaccine-mismatched mutations. To account for differential resolution of information from differing sequence counts per person, we use competing risks Cox regression with modeled causes of failure and propose an empirical Bayes approach for the classification model. Simulation studies demonstrate that our approach reduces bias, provides nominal confidence interval coverage, and improves statistical power compared to conventional methods. We apply our method to the HVTN 705 Imbokodo trial, which assessed the efficacy of a heterologous vaccine regimen in preventing HIV-1 infection.
\end{abstract}


\maketitle

\section{Introduction}

Developing vaccines is particularly challenging for pathogens with significant genetic diversity, such as HIV-1. While a vaccine may be effective in protecting against an infection or disease caused by certain pathogen genotypes, it may fail against others. Therefore, it is important for vaccine trials to assess the vaccine efficacy (VE) as a function of the breakthrough pathogen genotype. Methods designed to address this question constitute \textit{sieve analysis}, which draws an analogy between the vaccine and a sieve \citep{gilbert_statistical_1998, gilbert_sieve_2001}. The vaccine, like a sieve, acts as a barrier to pathogen infection, blocking certain virus types while allowing others to pass through and cause disease. A sieve effect can be evaluated by treating pathogen genotypes as competing risks, computing genotype-specific case rates in the vaccine and placebo groups, and contrasting these rates to estimate genotype-specific vaccine efficacy. We find evidence of a \textit{sieve effect} when VE varies across genotypes. 

When evaluating sieve effects, one must specify the pathogen
characteristics that may modify vaccine efficacy. These characteristics,
called \textit{marks}, are often defined by the amino acid divergence between
the infecting virus and the strain included in the vaccine construct. Marks may be binary or categorical, such as whether the strains are
matched at selected amino acid residues, or continuous, such as the
Hamming distance within a viral protein. Sieve effects are assessed by examining whether vaccine efficacy varies across mark values. Because
marks describe the breakthrough pathogen, they are observed only among
participants who experience the endpoint. Methods for sieve analysis
have been developed for different settings
\citep{gilbert_sieve_2001, sun_testing_2008,
juraska_mark-specific_2016, benkeser_estimating_2019}. Here, we focus on
binary marks.

Sieve analyses are particularly complex for viruses like HIV-1, since the viral population within an individual consists of an evolving collection of diverse variants, known as a \textit{quasispecies}. Due to a genetic bottleneck at transmission, a limited number of viral particles break through and may establish infection \citep{joseph_bottlenecks_2015}. Over time, immune pressures affect the evolution of the viral quasispecies within the individual. Historically, Sanger sequencing techniques detected only a small fraction of the viral variants present in an infection \citep{gregori_viral_2016}. Advances in sequencing technology, however, have enabled more comprehensive characterization of a quasispecies. With this technology, called \textit{deep sequencing}, it is possible to obtain hundreds of sequences from a viral sample, enabling detection of minor variants.

Although recent HIV-1 vaccine studies have used deep sequencing to characterize infecting viruses, analyses often reduce each participant's multi-sequence data to a single representative mark \citep{juraska_prevention_2024}, such as by selecting the modal mark value and then applying traditional single-sequence sieve methods. This approach can identify modal sieve effects but may miss what we refer to as \textit{tail} sieve effects. Table~\ref{tab:hypothetical_data} provides a toy example of such an effect. Although the modal marks are the same in the two treatment arms, the full sequence sets reveal a small number of vaccine-mismatched viruses in the vaccine arm and none in the placebo arm. Thus, the intra-individual mark distributions may differ between arms even when the modal marks do not. Examples like these motivate the need to define new estimands and develop estimation procedures capable of detecting these tail effects.

\begin{table}[htbp]
\centering
\begin{tabular}{clccc}
\hline
\multicolumn{2}{c}{} &
\multicolumn{1}{c}{\makecell{\textbf{Sanger sequencing}}} &
\multicolumn{2}{c}{\makecell{\textbf{Deep sequencing}}} \\
\cline{3-5}
\textbf{Subject ID} & \textbf{Arm} &
\makecell{\textbf{Sequence match (0)}\\ \textbf{or mismatch (1)}} &
\makecell{\textbf{Sequence set match (0)}\\ \textbf{or mismatch (1)}} &
\makecell{\textbf{Modal}\\ \textbf{sequence}} \\
\hline
1 & Placebo & 0 & 0, 0, 0, 0, 0 & 0 \\
2 & Placebo & 0 & 0, 0, 0, 0, 0 & 0 \\
3 & Placebo & 0 & 0, 0, 0, 0, 0 & 0 \\
4 & Placebo & 0 & 0, 0, 0, 0, 0 & 0 \\
5 & Placebo & 0 & 0, 0, 0, 0, 0 & 0 \\
6 & Vaccine & 0 & 0, 0, 0, 0, 1 & 0 \\
7 & Vaccine & 0 & 0, 0, 0, 0, 1 & 0 \\
8 & Vaccine & 0 & 0, 0, 0, 0, 1 & 0 \\
9 & Vaccine & 0 & 0, 0, 0, 0, 1 & 0 \\
10 & Vaccine & 0 & 0, 0, 0, 0, 1 & 0 \\
\hline
\end{tabular}
\caption{Toy dataset illustrating HIV infections in a hypothetical vaccine trial with five infections in each arm. For each individual, Sanger sequencing yields one (or occasionally a small handful of) sequence mark(s), coded as a match (0) or mismatch (1) to the vaccine strain; for simplicity, the table displays a single Sanger-derived mark per individual. In contrast, deep sequencing yields a set of sequence marks per individual. The Sanger-derived sequence marks are identical for infections in the vaccine and placebo arms. However, deep sequencing reveals a small minority of vaccine-mismatched marks that appear only in the vaccine-arm infections. This tail sieve effect is not detectable when only the modal sequence is used.}
\label{tab:hypothetical_data}
\end{table}

One important feature of deep sequencing data is the varying number of sequences for each sample, known as 
\textit{sequencing depth}. The sequencing depth determines the resolution of the sequencing data, and impacts the ability to accurately characterize the viral population within an individual. Sequencing depth can vary for several reasons, some potentially informative (e.g., differing viral load in samples), and ignoring this heterogeneity can bias analyses because samples with higher depth are more likely to detect rare variants, potentially creating spurious differences between groups \citep{garner_confounded_2011}. Therefore, it is imperative for any method with this data type to correct for the varying sequencing depth across individuals. 

We propose a novel approach to estimate the sieve effect for binary marks using deep sequencing data. We define an estimand for multi-sequence data that classifies failures by the presence or absence of the feature of interest in their sequence set, correcting for differing resolution due to sequencing depth. To estimate sieve effects, we employ a competing risks Cox model with a classified failure cause and use an empirical Bayes approach to define the classification model. Variation in sequencing depth can be viewed as a measurement error problem in which observed failure types are error-prone indicators of the true failure types. 

Our approach provides an alternative to the competing risks Cox regression method of \citet{van_rompaye_estimation_2012} for misclassified failure causes. Whereas their method assumes known and fixed misclassification probabilities, ours estimates participant-specific classification probabilities from sequencing and covariate information. Although empirical Bayes methods are widely used to address measurement error in econometrics \citep{walters_empirical_2024}, they have been less commonly applied in biostatistics and rarely in competing risks settings; see \citet{whittemore_errors--variables_1989} for a related example. We provide theoretical justification and demonstrate good performance in simulations. Related multi-sequence sieve methods include \citet{follmann_sieve_2018}, who studied estimands based on pathogen presence and count, and \citet{decamp_assessing_2013}, who compared generalized estimating equations with multiple outputation. Unlike these approaches, our method explicitly addresses measurement error caused by variable sequencing depth.

To ground the methodology, we illustrate our approach using deep sequencing data from the Imbokodo/HVTN 705 HIV-1 vaccine efficacy trial (NCT03060629). The trial enrolled females, aged 18–35 years, across five southern African countries and randomized participants 1:1 to receive either placebo or a vaccine regimen comprised of an Ad26-vector mosaic HIV-insert prime followed by a clade C gp140 protein. Although the study did not demonstrate significant efficacy against HIV-1 infection \citep{gray_mosaic_2024}, viral samples from participants who acquired HIV-1 were deep-sequenced using PacBio technology to characterize within-host diversity in the \textit{env} gene. These data provide an example where multiple sequences are available per individual, with variable sequencing depth across samples.

\section{Data structure and estimand}

For individual $i$, denote treatment assignment as $Z_i \in \{0,1\}$, stratum $S_i \in \{ 1,...,L\}$, and other covariates as a $p$-dimensional vector $X_i$. Define $W_i := (Z_i, X_i^\top)^\top$. Let $T_i$ be time from randomization until the study endpoint (first HIV-positive sample collection) and $C_i$ be time to right-censoring. Denote right-censored failure time $\tilde{T}_i := \min (T_i, C_i)$ and failure indicator $\Delta_i := I(T_i \le C_i)$. For participants who experience the study endpoint before censoring, i.e., $\Delta_i = 1$, multiple sequences of the virus are obtained with deep sequencing technology. For each sequence, we consider a binary feature taking values 0 or 1. We observe multiple instances of this feature per individual, which represents the feature's distribution within the viral quasispecies. One example of such a feature, used throughout the following sections, is whether a sequence matches (0) or mismatches (1) a specific residue in a given vaccine-insert virus. 

 We denote random variable $Q_i$ as the true proportion of sequences circulating in the blood that are mismatched to the vaccine. Note that this variable is not observed. Instead, we observe the proportion among $M_i$ sequences, where $M_i$ is an individual's sequence depth. We denote the binary match/mismatch mark for intra-individual sequences as $V_{i,1}, V_{i,2}, \ldots, V_{i,M_i}$, where each $V_{i,j} \in \{0, 1\}$ is observed only if $\Delta_i = 1$.

\begin{assumption}[Conditional simple random sampling of sequences]
For each individual $i$ with $\Delta_i=1$, conditional on
$(Q_i,M_i, W_i,S_i,\widetilde{T}_i)$, the observed sequence marks
$V_{i,1},V_{i,2},\ldots,V_{i,M_i}$ are independent and identically
distributed as $\operatorname{Bernoulli}(Q_i)$. Thus, once the underlying
mismatch proportion and sequencing depth are fixed, the remaining observed
variables provide no additional information about the sampled sequence marks.
\label{assumption:simple_random}
\end{assumption}

  Let $K_i := \sum_{j=1}^{M_i} V_{i,j}$ denote the total number of mismatched sequences observed for individual $i$. Under Assumption \ref{assumption:simple_random}, the conditional distribution of $K_i \mid M_i, Q_i \sim \text{Binomial}(M_i, Q_i)$. There are $n$ observations of the data, denoted as $\{W_i, S_i, \tilde{T}_i, \Delta_i, \Delta_i K_i, \Delta_i M_i \}_{i=1}^n$. 

 Similar to the $VE_{IF}$ estimand studied in \citet{follmann_sieve_2018}, we wish to measure vaccine efficacy against viral quasispecies with or without some presence of vaccine-mismatched viruses. Formally, we define the mark of interest as $J_{i,q_0}:=I(Q_i\geq q_0)$, indicating that the mismatch proportion is at least a small, fixed threshold $q_0$. There is no universal standard value for $q_0$; its choice should be guided by the scientific question, because different thresholds correspond to different definitions of meaningful feature presence. Although one scientific goal may be to distinguish viral quasispecies with any nonzero presence of the feature, finite sequencing depth prevents reliable classification using an exact zero threshold in practice. The implications of this resolution constraint and considerations for selecting $q_0$ are discussed in Web Appendix~C.
 
 We let $ \lambda_{js}(t; z, x)$ denote the covariate-adjusted conditional cause-specific hazard of disease for a viral quasispecies with mark $J_{q_0} = j$ for $j \in \{0,1\}$:
 \begin{equation*}
 \lambda_{js}(t; z, x) := \lim_{\delta \to 0} \frac{P(T \in [t, t + \delta), J_{q_0} = j\mid T \ge t, Z =z, X = x, S = s)}{\delta}
 \end{equation*}
 That is, $\lambda_{0s}(t; z, x)$ represents the hazard of disease at time $t$ caused by a viral quasispecies with less than a threshold $q_0$ of mismatched viruses for the $Z=z$ treatment arm with covariates $X = x$ in strata $S=s$, and $\lambda_{1s}(t; z, x)$ represents this cause-specific hazard for a quasispecies with mismatched viruses at least that threshold.
 Our estimand of interest is the vaccine efficacy against a viral quasispecies with mark $J_{q_0} = j$ for $j \in \{0,1\}$, denoted as $VE_j(t; x, s)$, which is defined as one minus the mark-specific covariate-adjusted hazard ratio comparing the vaccine and placebo arms:
 \begin{equation}
     VE_j(t;x, s) := 1 - \frac{ \lambda_{js}(t; 1, x)}{ \lambda_{js}(t; 0, x)}
\label{eq:ve_estimand}
 \end{equation}
 If the value $J_{i, q_0}$ were known for individuals with the study endpoint, we could use a competing risks Cox model to estimate \eqref{eq:ve_estimand}, where infections by viral quasispecies with and without the presence of vaccine-mismatched viruses ($J_{q_0} = 1$ and $J_{q_0} = 0$) are considered competing failure types \citep{prentice_analysis_1978, gilbert_comparison_2000}. However, for each individual $i$, $J_{i,q_0}$ is unknown because the true mismatch proportion $Q_i$ is unknown. We could use the observed empirical proportions $\tilde{Q}_i := \frac{K_i}{M_i}$ as a proxy for $Q_i$ and the empirical indicator $\tilde{J}_{i, q_0} := I(\frac{K_i}{M_i} \ge q_0)$ as a proxy for $J_{i,q_0}$. However, using this naïve proportion without any correction for measurement error can lead to highly biased results with loss of power in detecting a sieve effect, as shown in our simulation study in Section \ref{sec:simulation_study}. 

\section{Methodology}

\subsection{Competing risks Cox model with modeled failure cause}
\label{sec:cox_measurement_error}

To account for the fact that we do not observe $J_{i, q_0}$, we propose a new methodology for competing risks Cox regression with modeled failure cause. Our method will rely on creating a \textit{classification model} of $J_{i, q_0}$ based on its proxies, which include sequencing depth $M_i$ and the number of observed mismatches $K_i$.  We then incorporate these probabilities into our partial likelihood. In order to allow estimation and inference with a Cox model, we make a proportional hazards assumption and a non-informative right censoring assumption: 
\begin{assumption}[Proportional hazards]
Treatment assignment $Z$ and covariate vector $X$ have a proportional effect on the hazard for each viral quasispecies type in each stratum. Therefore, for virus type $J_{q_0} = j$ in stratum $s$, we assume
\begin{equation}
\label{eq:prop_hazards}
\begin{aligned}
    \lambda_{js}(t; z, x) &= \exp\big(\beta_j z + \alpha_j^\top x\big)\,\lambda_{0,js}(t),
\end{aligned}
\end{equation}
where $\lambda_{0,js}(t)$ is the stratum- and type-specific baseline hazard, and 
$\{\beta_0, \beta_1, \alpha_0, \alpha_1\}$ is the vector of regression parameters. To simplify notation, define $\theta_j := (\beta_j, \alpha_j^\top)^\top$. Then $
\lambda_{js}(t; w) = \exp(\theta_j^\top w)\,\lambda_{0,js}(t).$
\label{assumption:prop_hazards}
\end{assumption}
\begin{assumption}[Non-informative right censoring]
Censoring time is independent of failure time and the underlying mismatch proportion conditional on treatment assignment, covariates, and stratum, i.e., $C \perp (T,Q)\mid(W,S)$.
\label{assumption:censoring}
\end{assumption}

Under Assumption \ref{assumption:prop_hazards}, our estimand of interest can be written as
\begin{equation}
    VE_j(t; x, s) = 1 - e^{\beta_j}, \quad j = 0, 1.
\label{eq:identification_res}
\end{equation}
  Equation \eqref{eq:identification_res} does not depend on time $t$, covariate vector $x$, or strata $s$, so we drop these from the notation for $VE$ from this point forward (e.g. $VE_0$ and $VE_1$).  
 If $J_{i, q_0}$ were observed for each individual $i$, then we can use standard competing risks Cox methodology and construct separate log partial likelihoods, denoted as $\ell_j(\theta_j)$ for $j \in \{0, 1 \}$, from the conditional probabilities of an observed event of each type, given one such event was observed in the strata-specific risk set at that time: 
\begin{equation*}
\begin{aligned}
\ell_j(\theta_j)
= \sum_{i=1}^n \int_0^\tau 
\left[
   \theta_j^\top W_i
   - \log \left\{
       \sum_{l:\, S_\ell = S_i} 
       Y_l(t)\,\exp\!\big(\theta _j^\top W_l \big)
     \right\}
\right] dN_{ij}(t)
\end{aligned}
\end{equation*}
where $Y_i(t) := I(\tilde{T}_i \ge t)$ is the at-risk indicator for person $i$, $N_{ij}(t) := I(T_i \le t, \Delta_i = 1, J_{i,q_0} = j)$ is the cause-specific counting process, and $\tau$ denotes the end of the observation period (any value
greater than or equal to the largest observed event time).
We can differentiate to obtain the estimating function $U_j(\theta_j)$:
\begin{equation*}
U_j(\theta_j) \;=\; \sum_{i=1}^n \int_0^\tau 
\Big\{ W_i - \overline{W}_{j,S_i}(t;\theta_j) \Big\} \, dN_{ij}(t),
\end{equation*}
where $\overline{W}_{j,s}(t;\theta_j) 
:= \frac{\sum_{l: S_l = s} Y_l(t) \, W_l \, e^{\theta_j^\top W_l}}
       {\sum_{l: S_l = s} Y_l(t) \, e^{\theta_j^\top W_l}}.$

However, we need to adjust this estimating function to account for the fact that $ J_{i, q_0} $ is unobserved for each study endpoint. We define a modified estimating function $U^{adj}_j(\theta_j)$ using the mean score approach, replacing the unknown score term with its expected value given observed variables $W_i$ and $\tilde{T}_i$ along with auxiliary variables $K_i$ and $M_i$ \citep{pepe_auxiliary_1994}:
\begin{align}
U^{adj}_j(\theta_j) \;=\; \sum_{i=1}^n \int_0^\tau 
\Big\{ W_i - \overline{W}_{j,S_i}(t;\theta_j) \Big\} \, \nu_{q_0}(j; M_i, K_i, W_i, S_i, \tilde{T}_i) \, dN_{i}(t),
    \label{eq:modified_estimating_eq}
\end{align}
where  
$\nu_{q_0}(j; M, K, W, S, \tilde{T}) := P(J_{q_0} = j \mid M, K, W, S, \tilde{T}, \Delta = 1)$  
denotes the classification probabilities that $J_{q_0} = 1$ (i.e., $Q \ge q_0$) or $J_{q_0} = 0$ (i.e., $Q < q_0$) given the observed variables and $N_{i}(t) := I(T_i \le t, \Delta_i = 1)$ is the counting process for any failure type. Note that, for $\Delta_i = 1$,
$\mathbb{E}\!\left[dN_{ij}(t) \,\big|\, M_i, K_i, W_i, S_i, \tilde{T}_i \right]
= \nu_{q_0}\!\left(j; M_i, K_i, W_i, S_i, \tilde{T}_i\right)\, dN_i(t),$
so $U^{adj}_j(\theta_j)$ replaces the unobserved cause-specific counting process
$dN_{ij}(t)$ with its conditional expectation given the observed variables.

$U^{adj}_j(\theta_j)$ can be seen as a weighted estimating equation, where each event contributes to the equation for each failure type weighted by the probability that the event was that failure type.  If $\nu_{q_0}$ is known, then we can use standard Cox model theory to show consistency and derive asymptotic variance estimates under the usual regularity conditions 
\citep{andersen_coxs_1982}. However, $\nu_{q_0}$ will need to be estimated, and we will need to account for the uncertainty in its estimation in the downstream variance estimates.

\subsection{Classification model $\nu_{q_0}$}
\label{section:classification}

Since the classification probabilities $\nu_{q_0}(j; M_i, K_i, W_i, S_i, \tilde{T}_i)$ are not observed, we will need to estimate them using a model. In this section, we only need to consider estimation of $\nu_{q_0}(1; M_i, K_i, W_i, S_i, \tilde{T}_i)$, since  $\nu_{q_0}(0; M_i, K_i, W_i, S_i, \tilde{T}_i) = 1 - \nu_{q_0}(1; M_i, K_i, W_i, S_i, \tilde{T}_i)$. 
First, by definition, we note that
\begin{equation}
\begin{aligned}
    \nu_{q_0}(1; M, K, W, S, \tilde{T}) &= P(Q \ge q_0 \mid M, K, W, S, \tilde{T}, \Delta = 1) \\
    &= \int_{q_0}^{1} f_{Q|M, K}(q \mid M, K, W, S, \tilde{T}, \Delta = 1) \, dq
\label{eq:nu_model}
\end{aligned}
\end{equation}
where, with slight abuse of notation, $f_{Q|M, K}(q \mid M, K, W, S, \tilde{T}, \Delta = 1)$ denotes the conditional density of mismatch proportion $Q$, given observed variables $M$, $K$, $W$, $S$, and $\tilde{T}$ among individuals with $\Delta = 1$.
Using Bayes’ rule, we can express this density as
\begin{equation*}
\begin{aligned}
f_{Q|M,K}(q \mid M, K, W, S, \tilde{T}, \Delta = 1) 
&= \frac{f_{K \mid M, Q}(K \mid q, M, W, S, \tilde{T}, \Delta = 1) \cdot f_{Q|M}(q \mid M, W, S, \tilde{T}, \Delta = 1)}{f_{K|M}(K \mid M, W, S, \tilde{T}, \Delta = 1)}.
\label{eq:bayes_decomp}
\end{aligned}
\end{equation*}
Following from Assumption~\ref{assumption:simple_random}, we have that 
$f_{K \mid M, Q}(K \mid q, M, W, S, \tilde{T}, \Delta = 1)$
 corresponds to the probability mass function of a binomial distribution with parameters $M$ and $q$, which we denote as $f_{\text{binom}}(K; M, q) := \binom{M}{K} q^K (1 - q)^{M - K}$. In order to simplify the second term in the numerator $f_{Q|M}(q \mid M, W, S, \tilde{T}, \Delta = 1)$, we rely on one additional assumption.
 
 \begin{assumption}[Sequence depth conditional independence]
 Denote $B := (W, S, \tilde{T})$. The true mismatch proportion $Q$ is independent of $M$ conditional on $B$ among observed failures, i.e. $Q \perp M | B, \Delta = 1$.
     \label{assumption:seq_depth_indep}
 \end{assumption}

From Assumption \ref{assumption:seq_depth_indep}, we have that $f_{Q|M}(q|M, B, \Delta = 1) = f_Q(q|  B, \Delta = 1)$. Finally, we can rewrite the denominator as a normalizing constant:
\begin{align}
f_{Q|M, K}(q|M, K, B, \Delta = 1) &= \frac{f_{\text{binom}}(K; M, q) \cdot f_Q(q|B, \Delta = 1)}{ \int_0^1 f_{\text{binom}}(K; M, q)\cdot f_Q(q| B, \Delta = 1) dq}.
\label{eq:misclassify_model}
\end{align}
 The only part of the right hand side that is unknown is $f_Q(q|B, \Delta = 1)$, which represents the \textit{prior (or mixing) distribution} of $Q$ given $B$ for those with $\Delta = 1$. We use parametric binomial mixture model techniques (i.e., parametric deconvolution) to estimate this prior.
Indeed, the information we have on $K_i$ mismatches out of the $M_i$ sequences for each individual $i$ for whom $\Delta_i = 1$ can provide information on the true distribution of $Q| B, \Delta = 1$. Under equation \eqref{eq:misclassify_model}, this corresponds to an empirical Bayes approach, where posterior distributions are calculated using a prior distribution estimated from the data itself \citep{robbins_empirical_1956}.

To enable estimation and inference, we assume that the density of $Q|B=b, \Delta = 1$ arises from a known, correctly specified parametric model: 
\begin{assumption}[Parametric model]
The conditional distribution of $Q \mid B = b, \Delta = 1$ is correctly specified by a parametric family
$\mathcal{F} = \{ f_Q(\cdot; \gamma) : \gamma \in \Gamma \}$. That is, for every $b$ with $\Pr(\Delta = 1 \mid B = b) > 0$, there exists a bin-specific
parameter value $\gamma_b \in \Gamma$ such that $Q \mid (B = b, \Delta = 1) \sim f_Q(\cdot; \gamma_b).$
\label{assumption:parametric_dist}
\end{assumption}
While Assumption~\ref{assumption:parametric_dist} may appear restrictive, the spline-based modeling approach with regularization described in the following section allows considerable flexibility in estimating these distributions \citep{efron_empirical_2016}. If $B$ contains only discrete categorical variables, the density of $Q$ may be estimated separately across levels of $B$. If $B$ contains continuous variables, the distribution of $Q$ may instead be modeled as a function of $B$. For simplicity, in the following sections we assume that $B$ contains only discrete categorical variables and estimate the conditional density within each stratum.

In practice, some levels of $B$ may contain too few observations to estimate the distribution of $Q$ reliably. In such settings, stronger conditional independence assumptions may be used to reduce the conditioning set. For example, assuming $Q \perp (M,\widetilde{T}) \mid W,S,\Delta=1$ yields
$f_{Q\mid M}(q\mid M,W,S,\widetilde{T},\Delta=1)
=
f_Q(q\mid W,S,\Delta=1)$,
so that the conditional distribution of $Q$ need only be estimated across levels of $(W,S)$ rather than $(W,S,\widetilde{T})$. This and other alternatives to Assumption~\ref{assumption:seq_depth_indep} are discussed in Web Appendix~B.

\subsubsection{Binomial mixture models}
\label{section:deconv_problem}

We begin by describing the binomial mixture model setup in the context of our problem.  
Without loss of generality, fix $B = b$.  
Our goal is to estimate the conditional density of $Q \mid B = b, \Delta = 1$, where $Q$ has support on $[0, 1]$.  
Among the total of $n$ observations, suppose the first $n'$ correspond to this subgroup.  In the binomial mixture formulation, an unknown distribution of $Q \mid B = b, \Delta = 1$, denoted by $G_b$ with density $g_b(q)$,  
generates unobserved realizations $\{ Q_1, \dots, Q_{n'} \}$.  For each observation $i = 1, \dots, n'$, we observe pairs $(K_i, M_i)$ satisfying $K_i \mid M_i, Q_i \sim \text{Binomial}(M_i, Q_i).$ Thus, the distribution of observed data $K \mid M, B = b, \Delta = 1$ is a binomial mixture with probability mass function  
\begin{equation}
f(k \mid m) = \int_0^1 f_{\text{binom}}(k; m, q)\, g_b(q)\, dq.
\label{eq:marginal_likelihood}
\end{equation}

Our objective is to estimate mixing density $g_b$ from the observed data.  
Under Assumption~\ref{assumption:parametric_dist}, we assume the mixing density $g_b(q)$ belongs to a parametric family indexed by the parameter vector denoted as $\gamma_b$. Thus, from equation \eqref{eq:marginal_likelihood}, the marginal likelihood of a single observation  
$K_i \mid M_i, B_i = b, \Delta_i = 1$  can be written as $f(k_i \mid m_i; \gamma_b) 
= \int_{0}^{1} f_{\text{binom}}(k_i; m_i, q)\, g(q; \gamma_b)\, dq$. A maximum likelihood estimator (MLE) of $\gamma_b$ can be calculated as
\begin{equation}
\widehat{\gamma}_b
= \arg\max_{\gamma_b} \prod_{i=1}^{n'} f(K_i \mid M_i; \gamma_b)
= \arg\max_{\gamma_b} 
\sum_{i=1}^{n'} \log f(K_i \mid M_i; \gamma_b).
\label{eq:marginal_mle}
\end{equation}

Although parametric choices such as a Beta mixing distribution provide analytical simplicity, their restricted shapes can be inadequate in practice. We therefore utilize the approach of \citet{efron_empirical_2016}, which models the mixing density $g_b(q)$ using a low-dimensional exponential family representation based on spline basis functions.  
Specifically, $g(q; \gamma_b)$ is written as
$$
g(q; \gamma_b) = \exp\{ Q(q)^\top \gamma_b - \phi(\gamma_b) \},
$$
where $Q(q)$ is a vector of spline basis functions (e.g., natural splines on $[0,1]$) and $\phi(\gamma_b)$ is the log-normalizing constant ensuring integration to one. 
The number of spline basis functions controls the smoothness and flexibility of $g_b$. Point masses at $Q=0$ and $Q=1$ may also be included in the mixing model; in this case, the notation $g(q)\,dq$ is understood to include both the continuous component on $(0,1)$ and the discrete probability masses at the boundaries.

The estimator $\widehat{\gamma}_b$ is obtained via maximum likelihood following the general form in equation~\eqref{eq:marginal_mle}, substituting the spline-based model for $g_b$. To reduce variability of the estimator, a penalized approach can be utilized:
\begin{equation*}
\widehat{\gamma}_b
= \arg\max_{\gamma_b} 
\sum_{i=1}^{n'} \log f(K_i \mid M_i; \gamma_b) - c_0 \| \gamma_b \|^2
\end{equation*}
where $c_0 > 0$ controls the amount of regularization. 
This penalty shrinks the spline coefficients toward zero, effectively encouraging smoother estimates of $g_b(q)$. 
This regularization yields lower variance in $\widehat{\gamma}_b$ at the cost of bias but improves numerical stability of the resulting estimates \citep{efron_empirical_2016}.

\subsubsection{Estimation of classification probabilities}
\label{sec:classification_summary}

For each level $b$ of $B$, we estimate the mixing distribution of $Q$
among individuals with $\Delta=1$ by obtaining the
marginal likelihood estimator $\widehat{\gamma}_b$. Combining equations \eqref{eq:nu_model} and \eqref{eq:misclassify_model}, the estimated
classification probability for individual $i$ is then
\begin{equation}
\widehat{\nu}_{q_0}(1;M_i,K_i,B_i)
=
\frac{
\displaystyle
\int_{q_0}^{1}
f_{\mathrm{binom}}(K_i;M_i,q)
g(q;\widehat{\gamma}_{B_i})\,dq
}{
\displaystyle
\int_{0}^{1}
f_{\mathrm{binom}}(K_i;M_i,q)
g(q;\widehat{\gamma}_{B_i})\,dq
},
\label{eq:classification_prob}
\end{equation}
with $
\widehat{\nu}_{q_0}(0;M_i,K_i,B_i)
=
1-\widehat{\nu}_{q_0}(1;M_i,K_i,B_i).$ The estimated classification probabilities combine information from each individual's observed sequences with information borrowed from other endpoint cases in the same stratum of $B$. When sequencing depth is low, the individual's data provide limited information and the estimate relies more heavily on the fitted mixing distribution; when sequencing depth is high, it is driven more strongly by the individual's observed data. See Web Appendix~E for further discussion.

\subsection{Variance estimation}
\label{section:overall_procedure}

The proposed estimator can be viewed as a two-step M-estimator in which the mixing-distribution parameters are first estimated and then used to construct the classification probabilities in Equation~\eqref{eq:modified_estimating_eq}. Under correct specification and identifiability of the mixing model, together with standard smoothness and Cox-model regularity conditions, standard two-step M-estimation theory implies that $(\widehat{\theta}_0,\widehat{\theta}_1)$ are consistent and asymptotically normal; this structure parallels that considered by \citet{gao_semiparametric_2005}. To account for uncertainty from both estimation stages, we use a nonparametric bootstrap \citep{efron_bootstrap_1992}. Within each bootstrap sample, we re-estimate the mixing distribution and classification probabilities and then solve the modified Cox estimating equations to obtain $\widehat{\theta}_0^{*(k)}$ and $\widehat{\theta}_1^{*(k)}$. The empirical covariance matrix of the bootstrap replicates is used to estimate the sampling covariance of $(\widehat{\theta}_0,\widehat{\theta}_1)$ and to construct confidence intervals and hypothesis tests.

\subsection{Hypothesis testing}
\label{sec:hypothesis_testing}

In addition to the standard Cox regression hypothesis test for each risk type $j$ that type-specific vaccine efficacy equals zero (i.e., $\beta_j = 0$), we consider two additional hypothesis tests. The first test examines whether the vaccine confers any effect against either of the two viral types $J_{q_0} \in \{0, 1\}$. Specifically, we test the null hypothesis that vaccine efficacy is zero for \textit{both} risk types:
\begin{equation*}
\begin{aligned}
H_{A0}: VE_j = 0 \text{ for } j \in \{0,1\} \\
H_{A1}: VE_j \ne 0 \text{ for } j = 0 \text{ or } j = 1.
\end{aligned}
\end{equation*}
Equivalently, these null hypotheses can be expressed as $H_{A0}: \beta_0 = \beta_1 = 0$ versus $H_{A1}: \beta_0 \ne 0 \text{ or } \beta_1 \ne 0$.  

The second test evaluates whether vaccine efficacy varies across the two viral types, testing for a \textit{sieve effect} with respect to the binary mark. The null and alternative hypotheses are:
\begin{equation*}
\begin{aligned}
    H_{B0}&: VE_0 = VE_1 \\
    H_{B1}&: VE_0 \ne VE_1
\end{aligned}
\label{eq:hypothesis_test_b}
\end{equation*}
or equivalently, $H_{B0}: \beta_1 - \beta_0 = 0$ versus $H_{B1}: \beta_1 - \beta_0 \ne 0$. This test is analogous to the Lunn--McNeil test for equality of covariate effects in a competing-risks Cox analysis \citep{lunn_applying_1995} and serves as the primary test for detecting differential vaccine protection by viral type.  

To conduct each test, we first estimate the variance--covariance matrix  
$\widehat{\Sigma} = \widehat{\text{Cov}}(\widehat{\beta}_0, \widehat{\beta}_1)$, obtained empirically from bootstrap replicates of the fitted Cox model.  For the first test, we can apply a joint Wald test on the joint null hypothesis that $\beta_0 = 0$ and $\beta_1 = 0$. To do this, we compute the following test statistic: 
\begin{equation}
    W_{A0}
= 
\begin{pmatrix}
\widehat{\beta}_0 \\
\widehat{\beta}_1
\end{pmatrix}^{\!\top}
\widehat{\Sigma}^{-1}
\begin{pmatrix}
\widehat{\beta}_0 \\
\widehat{\beta}_1
\end{pmatrix}.
\end{equation}
We then obtain a p-value of $p = 1 - F_{\chi^2_2}(W_{A0})$, where $F_{\chi^2_2}$ is the cdf of a $\chi^2_2$ distribution with 2 degrees of freedom. 

For the second test, we can obtain the following Wald z-statistic: 
\begin{equation}
W_{B0} = \frac{\widehat{\beta}_1 - \widehat{\beta}_0}{\sqrt{\widehat{\mathrm{Var}}(\widehat{\beta}_1 - \widehat{\beta}_0)}}
\end{equation}
where $\sqrt{\widehat{Var}(\widehat{\beta}_1 - \widehat{\beta}_0)}$ is obtained by applying the delta method on the estimated variance-covariance matrix $\widehat{\Sigma}$. We then  obtain a two-sided p-value of $p = 2 \Phi (- |W_{B0}|)$ where $\Phi$ is the cdf of a standard Normal distribution.

\section{Simulation study}
\label{sec:simulation_study}

We study the performance of our proposed method with a numerical study. First, we fix threshold $q_0 = 0.01$. We consider two equal-sized treatment arms: placebo ($Z = 0$) and vaccine ($Z = 1$), with sample sizes in each arm of 1,000, 2,000, and 3,000. We assume a single stratum for all individuals and simulate a binary adjustment covariate $X$ for each individual. We set the probability of the event by $t=5$ in the placebo arm at roughly 15\%. We consider three vaccine efficacy scenarios:
\begin{itemize}
    \item Setting (a) (no VE): 0\% VE for both viral types.
    \item Setting (b) (positive VE, no sieve effect): 50\% VE against both viral types. 
    \item Setting (c) (positive VE with sieve effect):  50\% VE against viral type $J_{q_0} = 0$ and 5\% VE against viral type $J_{q_0} = 1$. 
\end{itemize}

Details of these exact simulation conditions are shown in Web Appendix F.1. We perform three simulation studies, each with settings (a)--(c), exploring different conditions for simulating sequencing depth $M$, with $K$ generated as $K|M, Q \sim \text{Binomial}(M, Q)$:

\begin{itemize}
    \item \textbf{Simulation Study \#1:} Large, fixed sequencing depth per case with $M = 2000$.
    \item \textbf{Simulation Study \#2:} Varying sequencing depth  but with equal variation per arm, with $M$ ranging uniformly from 1 to 15 in 40\% of endpoint cases, and uniformly from 16 to 1,000 in the remaining endpoint cases. 
    \item \textbf{Simulation Study \#3:} Unequal sequencing depth across arms. In the placebo arm, $M$ ranges uniformly from 1 to 15 in 20\% of endpoint cases and 16 to 1,000 in the remaining cases. In the vaccine arm, $M$ ranges uniformly from 1 to 15 in 40\% of endpoint cases and 16 to 1,000 in the remaining cases.
\end{itemize}
Across all conditions, we compared bias, standard error, and confidence interval coverage for the proposed and uncorrected estimators over 1,000 simulations. The uncorrected estimator used the empirical failure indicator $\widetilde{J}_{i,q_0}=I(K_i/M_i\ge q_0)$. For the proposed estimator, we assumed the stronger condition $Q\perp(M,\widetilde{T})\mid Z,X,\Delta=1$, which holds under the data-generating mechanism. Within each level of $(Z,X)$, we estimated the mixing distribution of $Q$ using penalized splines with $df=5$ and $c_0=0.01$ \citep{narasimhan_deconvolver_2020}, augmented with point masses at 0 and 1. Although the spline models are not correctly specified, they provide a flexible and more realistic evaluation than correctly specified parametric models.

Standard errors were estimated using 1,000 bootstrap samples, and Wald confidence intervals were constructed accordingly. We tested for a sieve effect using $H_{B0}$ in Section~\ref{sec:hypothesis_testing}. For the uncorrected estimator, we fit a standard competing-risks Cox model and used the Lunn--McNeil test for equal covariate effects \citep{lunn_applying_1995}. In settings (a) and (b), where $VE_0=VE_1$, we compared Type~I error rates; in setting (c), where $VE_0\ne VE_1$, we compared power. Web Appendix~F presents additional simulations with correctly specified mixing distributions and varying choices of $q_0$.
\subsection{Simulation Study \#1}
Results for Simulation Study \#1 are presented in Web Appendix F.2. Both the proposed and uncorrected estimators perform well, as expected, because misclassification error is rare with large sequencing depths.

\subsection{Simulation Study \#2}

\begin{figure}[h!]
    \centering
    \includegraphics[width=\textwidth]{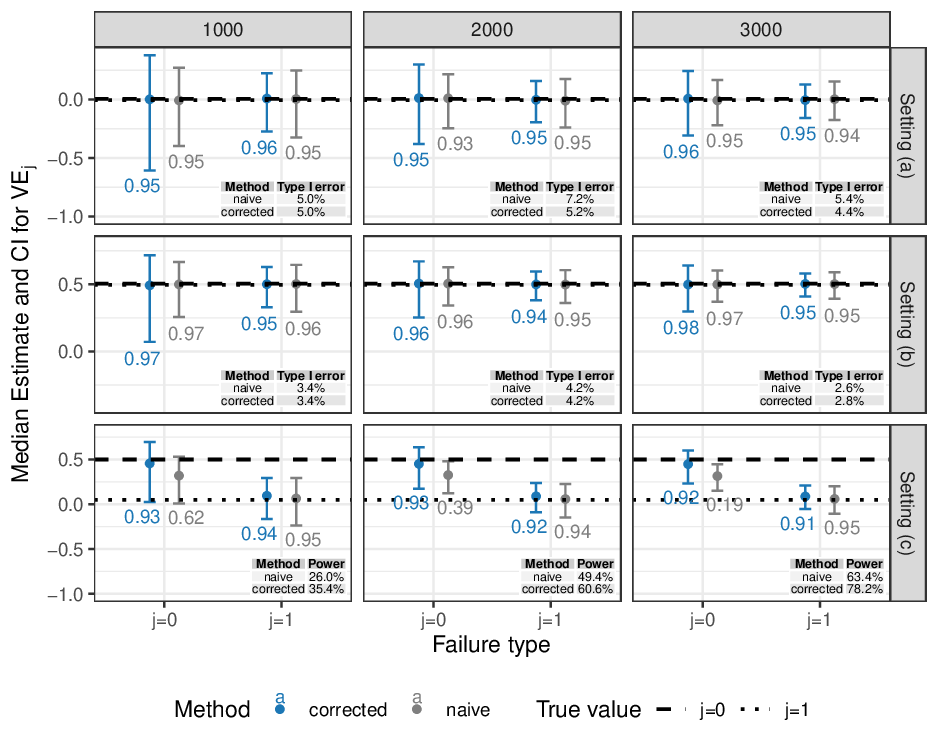}
    \caption{Results for Simulation Study \#2. The top, middle, and bottom panels display results for settings (a), (b), and (c) respectively. The left, middle, and right panels display results for differing sample sizes per treatment arm. Each graph displays points for median $VE$ estimates for each failure type across all simulations, along with median lower and upper 95\% confidence interval bounds. The dashed and dotted horizontal lines are placed at the true values for the $j=0$ and $j=1$ failure types, respectively. The numbers below each error bar display the confidence interval coverage. We compare Type I error between the two estimators in settings (a) and (b) and power to detect a sieve effect in setting (c).}
    \label{fig:sim_fig2}
\end{figure}

Results for Simulation Study~\#2 are shown in
Figure~\ref{fig:sim_fig2}. When sequencing depth varied similarly
between treatment arms, both estimators performed well in settings (a)
and (b), where vaccine efficacy was equal across failure types.
Although misclassification occurred primarily from $j=1$ to $j=0$, it
was approximately balanced between arms, yielding minimal bias,
near-nominal coverage, and controlled Type~I error. In setting (c),
however, the uncorrected estimator was substantially biased for the
$j=0$ failure type and exhibited confidence interval undercoverage,
whereas the proposed estimator remained approximately unbiased,
maintained near-nominal coverage, and had greater power to detect the
sieve effect.

\subsection{Simulation Study \#3}

\begin{figure}[h!]
    \centering
    \includegraphics[width=\textwidth]{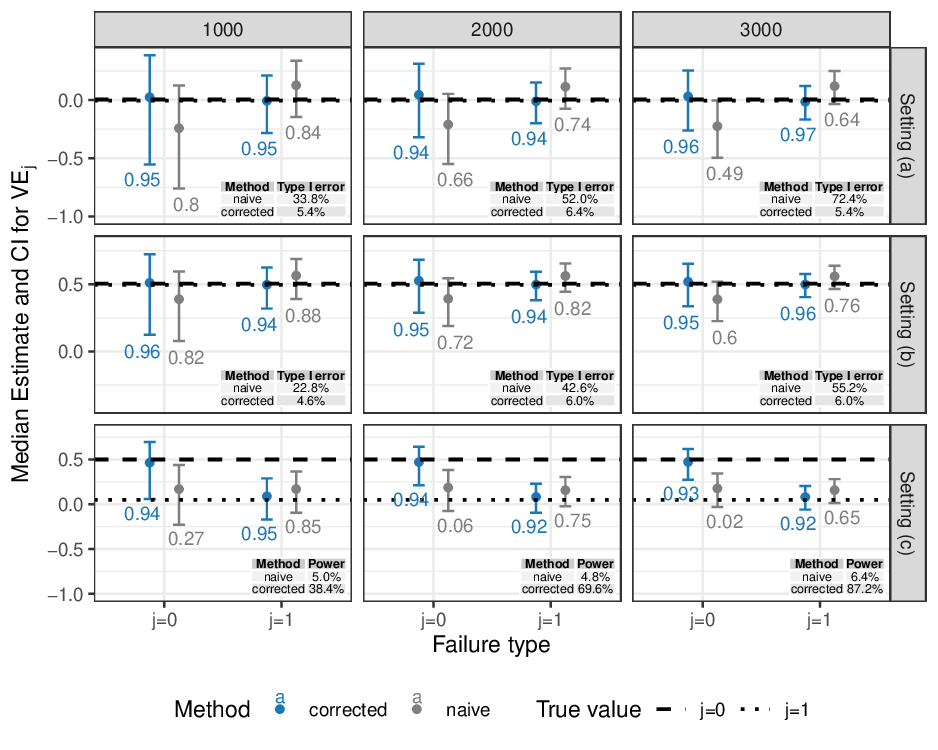}
    \caption{Results for Simulation Study \#3. The top, middle, and bottom panels display results for settings (a), (b), and (c) respectively. The left, middle, and right panels display results for differing sample sizes per treatment arm. Each graph displays points for median $VE$ estimates for each failure type across all simulations, along with median lower and upper 95\% confidence interval bounds. The dashed and dotted horizontal lines are placed at the true values for the $j=0$ and $j=1$ failure types, respectively. The numbers below each error bar display the confidence interval coverage. We compare Type~I error between the two estimators in settings (a) and (b) and power to detect a sieve effect in setting (c).}
    \label{fig:sim_fig3}
\end{figure}

Results for Simulation Study~\#3 are shown in
Figure~\ref{fig:sim_fig3}. When sequencing depths differed
between treatment arms, differential misclassification led to biased
vaccine efficacy estimates and substantial Type~I error inflation for
the uncorrected estimator in settings (a) and (b), with inflation
increasing with sample size. The proposed estimator provided much
better Type~I error control, although very mild Type~I error inflation remained because
the spline models did not perfectly represent the true mixing
distributions. In setting (c), the uncorrected estimator had power close
to 5\% across all sample sizes, whereas power for the proposed estimator
increased from 38.4\% to 87.2\% as the sample size increased from 1,000
to 3,000 participants per arm.

Additional simulations in Web Appendix F showed similar performance under correctly specified mixing models and across varying choices of $q_0$, with the proposed estimator maintaining lower bias and better Type~I error control and power than the uncorrected estimator.

\section{Data application}
\label{sec:data_application}
We applied our proposed methodology to the HVTN 705 HIV-1 vaccine efficacy trial \citep{gray_mosaic_2024, juraska_quantifying_2026}. Two classes of binary sequence features were analyzed: (1) reference-dependent marks indicating whether the amino acid residue in a given sequence position matches or mismatches that in each of three HIV-1 Env vaccine insert sequences (Mos1 and Mos2S in the Ad26-vector prime, and C97ZA in the Clade C boost), and (2) reference-independent amino acid indicators denoting the presence of a specific amino acid at a given position (using the HXB2 numbering system). The median sequencing depth across all samples was 104 (IQR: 21.5--200.5), with lower depths in the vaccine arm (median [IQR]: 84 [21--147.25]) than in the placebo arm (127 [26.5--259]). Based on these depths, following the guidance in Web Appendix C, we set thresholds of 1\% and 99\%. Therefore, for each feature, the resulting analyses estimate vaccine efficacy against quasispecies with $<1\%$ (or $<99\%$) presence of the feature compared with those with $\ge 1\%$ (or $\ge 99\%$) presence. Age, BMI, and HIV risk score were included as adjustment covariates $X$, with stratification ($S$) by the indicator of South Africa. The mixing distributions in the classification model (conditioning on vaccination status only) were estimated using splines ($df = 10, c_0 = 1$). For comparison with conventional sieve analyses, we also report results based on the modal sequence mark. Although this benchmark targets a different estimand, it illustrates the conclusions that would be obtained from a single-sequence analysis.

\begin{figure}[!htbp]
    \centering
    \includegraphics[width=1\textwidth]{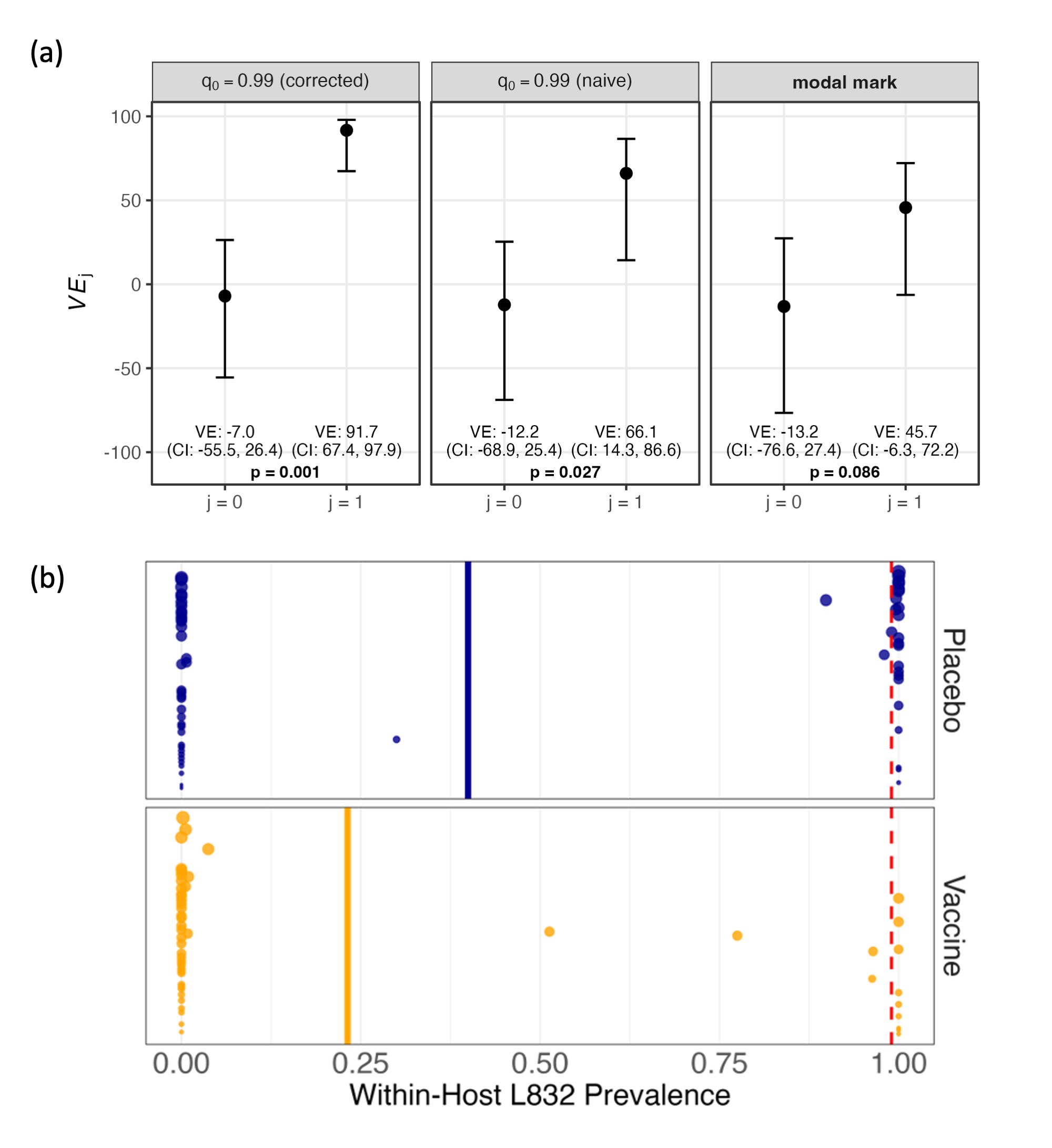}
    \caption{(a) Results for the sieve analysis of leucine at HXB2 position 832 comparing the proposed estimator, uncorrected estimator using a raw $\ge 99\%$ classification and uncorrected estimator using the modal mark (i.e., raw $\ge 50\%$ classification). (b) For each individual, the plot shows the raw proportion of the individual's sequences with leucine at HXB2 position 832. The size of the dot is proportional to the sequencing depth for that individual. The dots are ordered vertically by sequencing depth. The solid lines show the average proportion for each arm.}
    \label{fig:application_fig}
\end{figure}

Because the data had thousands of potential sequence features, we perform a pre-screening procedure to reduce the starting sequence set, described in Web Appendix D. Among the 8,458 candidate binary features, 114 feature-threshold combinations passed screening: 71 at the 1\% threshold and 43 at the 99\% threshold. We identified evidence of differential VE for amino acid leucine at HXB2 position 832 in the envelope protein under the 99\% threshold (unadjusted $p<0.001$; FWER-adjusted $p=0.023$). Quasispecies with $\ge 99\%$ leucine at this position exhibited high vaccine efficacy (VE estimate: 91.7\%; 95\% CI: 67.4, 97.9), whereas those with $<99\%$ leucine showed no evidence of protection (VE estimate: $-7.0\%$; 95\% CI: -55.5, 26.4) (Figure \ref{fig:application_fig}a). Results from a single-sequence sieve analysis using the naïve $\ge 99\%$ classification (unadjusted $p = 0.027$) and the modal sequence mark (unadjusted $p = 0.086$) showed a similar trend in differential $VE$ but had higher p-values. Figure \ref{fig:application_fig}b shows the raw data for this viral feature. As shown, the vaccine arm had a lower raw proportion of viral quasispecies exceeding the 99\% leucine threshold at this position, and cases in the vaccine arm meeting this threshold tended to have lower sequencing depths. Our method corrects for the possible misclassification into the $\ge 99\%$ bin due to low sequencing depth.

This result suggests that the presence of a minority of variants with a non-leucine residue at that position may be associated with reduced vaccine protection. One possibility is that substitutions at this position indirectly alter the amount or conformation of surface-expressed Env, thereby influencing humoral recognition of infected cells. Alternatively, the lower prevalence of L832 among vaccine recipients may reflect indirect or post-infection adaptation involving the LLP-1 motif, although this remains speculative, particularly given that position 832 was not included in the vaccine immunogens.

\section{Discussion}

Our proposed methodology provides a principled framework for evaluating sieve effects in prevention trials that have deep viral sequencing data. The approach is particularly well suited for rapidly evolving viruses that generate substantial within-host diversity, such as HIV, where analyses based solely on a modal or mindist sequence may miss important intra-individual variation. We believe our work has two main innovations in the context of sieve analysis literature: (1) the definition of an estimand in the context of deep sequencing data, and (2) an estimation procedure that explicitly addresses heterogeneous sequencing depth, which is a key feature of the data. In contrast with naïve approaches that ignore measurement error or exclude participants with low sequencing depth, our method retains all available information and explicitly corrects for sequence depth-related measurement error.

Our framework addresses settings in which the true failure type is not directly observed and must instead be inferred from error-prone sequencing data. Unlike prior methods that assume known and fixed misclassification probabilities, our approach allows these probabilities to vary across cases and estimates them from the sequencing data. The empirical Bayes posterior combines individual-level information from the number of mismatches $K_i$ and sequencing depth $M_i$ with information from the estimated mixing distribution. At high sequencing depth, classification relies primarily on the individual's observed sequences; at low sequencing depth, it shrinks toward the population distribution. More broadly, the method illustrates how empirical Bayes approaches can combine subject-specific and population-level information to address measurement error in nontraditional data settings. Extensions of the proposed methodology to more than two failure types and to settings with missing sequencing data are described in Web Appendix A.

One limitation of our proposed approach concerns the modeling assumptions required to construct the classification model. The method requires estimation of the binomial mixing distribution of $Q$ conditional on some combination of $(W,S,\widetilde{T})$. When the covariate vector is high-dimensional, estimation of this conditional distribution may become difficult and may limit the scalability of the approach. One possible simplification is to reduce the conditioning set used in the classification model. However, smaller conditioning sets require stronger conditional independence assumptions, and these assumptions require appropriate scientific justification; see Web Appendix~B for further discussion. Finally, valid inference depends on correct specification of the mixing distributions of $Q$. Penalized splines provide a flexible modeling strategy and performed well in our simulation studies, but misspecification may nevertheless lead to biased estimates and invalid confidence intervals or $p$-values.

Another limitation is that our method uses threshold-based dichotomization of the underlying viral features, reflecting the empirical reality that the observed proportions are often concentrated near 0 or 1. An alternative approach would treat these proportions as continuous features and incorporate a continuous measurement error model to account for variable sequencing depth. 

\backmatter

\section*{Acknowledgements}
We thank the reviewer and Associate Editor for their helpful comments; Craig Magaret for constructing the data set; Elena Giorgi, Paul Edlefsen, Raabya Rossenkhan, Allan deCamp, and James Ludwig for helpful discussions; and the participants and investigators of HVTN 705. Research reported in this publication was supported by the National Institutes of Health (NIH) under award number R37AI054165, and the National Institute of Allergy and Infectious Diseases of the NIH under the U.S. Public Health Service Grant AI068635. 

\section*{Supplementary Material}
Supplementary materials, including Web Appendices A–F and additional figures and tables, are provided following the main manuscript. R code for the simulations and data analysis is also available at \href{https://github.com/jpspeng/multiseq_cox_sieve.git}{https://github.com/jpspeng/multiseq\_cox\_sieve.git}.

\section*{Data Availability} 
For the HVTN 705 data application, the data sharing policy of Johnson \& Johnson is available at \url{https://innovativemedicine.jnj.com/our-innovation/clinical-trials/transparency}. As noted on this site, requests for access to the study data can be submitted through the Yale Open Data Access (YODA) Project at \url{http://yoda.yale.edu}.

\clearpage

\bibliographystyle{biom}
\bibliography{references}

@article{van_rompaye_estimation_2012,
	title = {{Estimation with {Cox} models: cause-specific survival analysis with misclassified cause of failure}},
	volume = {23},
	issn = {1044-3983},
	shorttitle = {Estimation with {Cox} models},
	url = {https://www.ncbi.nlm.nih.gov/pmc/articles/PMC3903130/},
	doi = {10.1097/EDE.0b013e3182454cad},
	number = {2},
	urldate = {2025-01-13},
	journal = {Epidemiology},
	author = {Van Rompaye, B and Jaffar, S and Goetghebeur, E},
	month = mar,
	year = {2012},
	pmid = {22317803},
	pmcid = {PMC3903130},
	pages = {194--202},
}

@article{efron_empirical_2016,
	title = {{Empirical {Bayes} deconvolution estimates}},
	volume = {103},
	issn = {0006-3444},
	url = {https://www.jstor.org/stable/43908598},
	number = {1},
	urldate = {2025-01-15},
	journal = {Biometrika},
	author = {Efron, Bradley},
	year = {2016},
	note = {},
	pages = {1--20},
}

@article{follmann_sieve_2018,
	title = {{Sieve {analysis} {using} the {number} of {infecting} {pathogens}}},
	volume = {74},
	issn = {0006-341X},
	url = {https://www.ncbi.nlm.nih.gov/pmc/articles/PMC6004265/},
	doi = {10.1111/biom.12833},
	number = {3},
	urldate = {2025-01-23},
	journal = {Biometrics},
	author = {Follmann, Dean and Huang, Chiung-Yu},
	month = sep,
	year = {2018},
	pmid = {29238965},
	pmcid = {PMC6004265},
	pages = {1023--1033},
}

@article{gilbert_comparison_2000,
	title = {{Comparison of competing risks failure time methods and time-independent methods for assessing strain variations in vaccine protection}},
	volume = {19},
	issn = {0277-6715},
	doi = {10.1002/1097-0258(20001130)19:22<3065::aid-sim600>3.0.co;2-d},
	language = {eng},
	number = {22},
	journal = {Statistics in Medicine},
	author = {Gilbert, Peter},
	month = nov,
	year = {2000},
	pmid = {11113943},
	pages = {3065--3086},
}

@article{juraska_prevention_2024,
	title = {{Prevention efficacy of the broadly neutralizing antibody {VRC01} depends on {HIV}-1 envelope sequence features}},
	volume = {121},
	issn = {0027-8424},
	url = {https://www.ncbi.nlm.nih.gov/pmc/articles/PMC10823214/},
	doi = {10.1073/pnas.2308942121},
	number = {4},
	urldate = {2025-01-23},
	journal = {Proceedings of the National Academy of Sciences of the United States of America},
	author = {Juraska, Michal and Bai, Hongjun and deCamp, Allan C. and Magaret, Craig A. and Li, Li and Gillespie, Kevin and Carpp, Lindsay N. and Giorgi, Elena E. and Ludwig, James and Molitor, Cindy and Hudson, Aaron and Williamson, Brian D. and Espy, Nicole and Simpkins, Brian and Rudnicki, Erika and Shao, Danica and Rossenkhan, Raabya and Edlefsen, Paul T. and Westfall, Dylan H. and Deng, Wenjie and Chen, Lennie and Zhao, Hong and Bhattacharya, Tanmoy and Pankow, Alec and Murrell, Ben and Yssel, Anna and Matten, David and York, Talita and Beaume, Nicolas and Gwashu-Nyangiwe, Asanda and Ndabambi, Nonkululeko and Thebus, Ruwayhida and Karuna, Shelly T. and Morris, Lynn and Montefiori, David C. and Hural, John A. and Cohen, Myron S. and Corey, Lawrence and Rolland, Morgane and Gilbert, Peter B. and Williamson, Carolyn and Mullins, James I.},
	year = {2024},
	pmid = {38241441},
	pmcid = {PMC10823214},
	pages = {e2308942121},
}

@incollection{robbins_empirical_1956,
	title = {{An {empirical} {Bayes} {approach} to {statistics}}},
	url = {https://projecteuclid.org/ebooks/berkeley-symposium-on-mathematical-statistics-and-probability/Proceedings-of-the-Third-Berkeley-Symposium-on-Mathematical-Statistics-and/chapter/An-Empirical-Bayes-Approach-to-Statistics/bsmsp/1200501653},
	urldate = {2025-01-23},
    booktitle = {Proceedings of the Third Berkeley Symposium on Mathematical Statistics and Probability},
publisher = {},
	author = {Robbins, Herbert},
	month = jan,
	year = {1956},
}

@article{gilbert_statistical_1998,
	title = {{Statistical methods for assessing differential vaccine protection against human immunodeficiency virus types}},
	volume = {54},
	issn = {0006-341X},
	language = {eng},
	number = {3},
	journal = {Biometrics},
	author = {Gilbert, Peter and Self, S. G. and Ashby, M. A.},
	month = sep,
	year = {1998},
	pmid = {9750238},
	pages = {799--814},
}

@article{gilbert_sieve_2001,
	title = {{Sieve analysis: methods for assessing from vaccine trial data how vaccine efficacy varies with genotypic and phenotypic pathogen variation}},
	volume = {54},
	issn = {0895-4356},
	shorttitle = {Sieve analysis},
	doi = {10.1016/s0895-4356(00)00258-4},
	language = {eng},
	number = {1},
	journal = {Journal of Clinical Epidemiology},
	author = {Gilbert, Peter and Self, S. and Rao, M. and Naficy, A. and Clemens, J.},
	month = jan,
	year = {2001},
	pmid = {11165470},
	pages = {68--85},
}

@article{juraska_mark-specific_2016,
	title = {{Mark-{specific} {hazard} {ratio} {model} with {missing} {multivariate} {marks}}},
	volume = {22},
	issn = {1380-7870},
	url = {https://www.ncbi.nlm.nih.gov/pmc/articles/PMC4848257/},
	doi = {10.1007/s10985-015-9353-9},
	number = {4},
	urldate = {2025-01-24},
	journal = {Lifetime Data Analysis},
	author = {Juraska, Michal and Gilbert, Peter B.},
	month = oct,
	year = {2016},
	pmid = {26511033},
	pmcid = {PMC4848257},
	pages = {606--625},
}

@article{whittemore_errors--variables_1989,
	title = {{Errors-in-{variables} {regression} {using} {Stein} {estimates}}},
	volume = {43},
	issn = {0003-1305},
	url = {https://www.jstor.org/stable/2685366},
	doi = {10.2307/2685366},
	number = {4},
	urldate = {2025-01-29},
	journal = {The American Statistician},
	author = {Whittemore, Alice S.},
	year = {1989},
	note = {},
	pages = {226--228},
}

@article{benkeser_estimating_2019,
	title = {{Estimating and {testing} {vaccine} {sieve} {effects} {using} {machine} {learning}}},
	volume = {114},
	issn = {0162-1459},
	doi = {10.1080/01621459.2018.1529594},
	language = {eng},
	number = {527},
	journal = {Journal of the American Statistical Association},
	author = {Benkeser, David and Gilbert, Peter and Carone, Marco},
	year = {2019},
	pmid = {31649413},
	pmcid = {PMC6812562},
	pages = {1038--1049},
}

@article{prentice_analysis_1978,
	title = {{The analysis of failure times in the presence of competing risks}},
	volume = {34},
	issn = {0006-341X},
	language = {eng},
	number = {4},
	journal = {Biometrics},
	author = {Prentice, R. L. and Kalbfleisch, J. D. and Peterson, A. V. and Flournoy, N. and Farewell, V. T. and Breslow, N. E.},
	month = dec,
	year = {1978},
	pmid = {373811},
	pages = {541--554},
}

@article{garner_confounded_2011,
	title = {{Confounded by sequencing depth in association studies of rare alleles}},
	volume = {35},
	issn = {1098-2272},
	doi = {10.1002/gepi.20574},
	language = {eng},
	number = {4},
	journal = {Genetic Epidemiology},
	author = {Garner, Chad},
	month = may,
	year = {2011},
	pmid = {21328616},
	pmcid = {PMC3129358},
	pages = {261--268},
}

@article{gregori_viral_2016,
	title = {{Viral quasispecies complexity measures}},
	volume = {493},
	issn = {1096-0341},
	doi = {10.1016/j.virol.2016.03.017},
	language = {eng},
	journal = {Virology},
	author = {Gregori, Josep and Perales, Celia and Rodriguez-Frias, Francisco and Esteban, Juan I. and Quer, Josep and Domingo, Esteban},
	month = jun,
	year = {2016},
	pmid = {27060566},
	pages = {227--237},
}

@article{joseph_bottlenecks_2015,
	title = {{Bottlenecks in {HIV}-1 transmission: insights from the study of founder viruses}},
	volume = {13},
	copyright = {2015 Springer Nature Limited},
	issn = {1740-1534},
	shorttitle = {Bottlenecks in {HIV}-1 transmission},
	url = {https://www.nature.com/articles/nrmicro3471},
	doi = {10.1038/nrmicro3471},
	language = {en},
	number = {7},
	urldate = {2025-06-22},
	journal = {Nature Reviews Microbiology},
	author = {Joseph, Sarah B. and Swanstrom, Ronald and Kashuba, Angela D. M. and Cohen, Myron S.},
	month = jul,
	year = {2015},
	note = {},
	pages = {414--425},
}

@article{decamp_assessing_2013,
	title = {{Assessing vaccine effects in {HIV}-1 vaccine trials: antigenic maps, antigen selection, and sieve analysis}},
	shorttitle = {Assessing vaccine effects in {HIV}-1 vaccine trials},
	url = {http://hdl.handle.net/1773/25123},
	language = {eng},
	urldate = {2025-06-22},
	journal = {University of Washington Libraries},
	author = {DeCamp, Allan},
	collaborator = {Gilbert, Peter},
	year = {2013},
	note = {OCLC: 882504213},
}

@misc{walters_empirical_2024,
	type = {Working {Paper}},
	series = {Working {Paper} {Series}},
	title = {{Empirical {Bayes} {methods} in {labor} {economics}}},
	url = {https://www.nber.org/papers/w33091},
	doi = {10.3386/w33091},
	urldate = {2025-06-22},
	publisher = {National Bureau of Economic Research},
	author = {Walters, Christopher R.},
	month = oct,
	year = {2024},
	doi = {10.3386/w33091},
}

@article{pepe_auxiliary_1994,
	title = {{Auxiliary outcome data and the mean score method}},
	volume = {42},
	issn = {0378-3758},
	url = {https://www.sciencedirect.com/science/article/pii/0378375894901945},
	doi = {10.1016/0378-3758(94)90194-5},
	number = {1},
	urldate = {2025-06-22},
	journal = {Journal of Statistical Planning and Inference},
	author = {Pepe, Margaret Sullivan and Reilly, Marie and Fleming, Thomas R.},
	month = nov,
	year = {1994},
	pages = {137--160},
}

@article{sun_testing_2008,
	title = {{Testing and estimation of time-varying cause-specific hazard ratios with covariate adjustment}},
	volume = {64},
	issn = {1541-0420},
	doi = {10.1111/j.1541-0420.2008.01012.x},
	language = {eng},
	number = {4},
	journal = {Biometrics},
	author = {Sun, Yanqing and Hyun, Seunggeun and Gilbert, Peter},
	month = dec,
	year = {2008},
	pmid = {18355384},
	pmcid = {PMC9841889},
	pages = {1070--1079},
}

@article{andersen_coxs_1982,
	title = {{Cox's {regression} {model} for {counting} {processes}: {a} {large} {sample} {study}}},
	volume = {10},
	issn = {0090-5364, 2168-8966},
	shorttitle = {Cox's {regression} {model} for {counting} {processes}},
	url = {https://projecteuclid.org/journals/annals-of-statistics/volume-10/issue-4/Coxs-Regression-Model-for-Counting-Processes--A-Large-Sample/10.1214/aos/1176345976.full},
	doi = {10.1214/aos/1176345976},
	number = {4},
	urldate = {2025-06-22},
	journal = {The Annals of Statistics},
	author = {Andersen, P. K. and Gill, R. D.},
	month = dec,
	year = {1982},
	note = {},
	pages = {1100--1120},
}

@article{gray_mosaic_2024,
  author  = {Gray, Glenda E. and Mngadi, Kathryn and Lavreys, Ludo and Nijs, Steven
             and Gilbert, Peter B. and Hural, John and Hyrien, Ollivier and Juraska, Michal
             and Luedtke, Alex and Mann, Philipp and McElrath, M. Juliana and others},
  title   = {{Mosaic {HIV}-1 vaccine regimen in southern {African} women ({Imbokodo}/{HVTN} 705/{HPX2008}): a randomised, double-blind, placebo-controlled, phase 2b trial}},
  journal = {The Lancet Infectious Diseases},
  year    = {2024},
  volume  = {24},
  number  = {11},
  pages   = {1201--1212},
  doi     = {10.1016/S1473-3099(24)00358-X}
}

@incollection{efron_bootstrap_1992,
	title = {{Bootstrap {methods}: {another} {look} at the {jackknife}}},
	isbn = {978-1-4612-4380-9},
	shorttitle = {Bootstrap {methods}},
	url = {https://link.springer.com/chapter/10.1007/978-1-4612-4380-9_41},
	language = {en},
	urldate = {2025-07-25},
	booktitle = {Breakthroughs in {Statistics}},
	publisher = {Springer, New York, NY},
	author = {Efron, Bradley},
	year = {1992},
	doi = {10.1007/978-1-4612-4380-9_41},
	pages = {569--593},
}

@article{lunn_applying_1995,
	title = {{Applying {Cox} {regression} to {competing} {risks}}},
	volume = {51},
	issn = {0006-341X},
	url = {https://www.jstor.org/stable/2532940},
	doi = {10.2307/2532940},
	number = {2},
	urldate = {2025-11-04},
	journal = {Biometrics},
	author = {Lunn, Mary and McNeil, Don},
	year = {1995},
	note = {},
	pages = {524--532},
}

@article{tarone_modified_1990,
	title = {{A modified {Bonferroni} method for discrete data}},
	volume = {46},
	issn = {0006-341X},
	language = {eng},
	number = {2},
	journal = {Biometrics},
	author = {Tarone, R. E.},
	month = jun,
	year = {1990},
	pmid = {2364136},
	pages = {515--522},
}

@article{gao_semiparametric_2005,
	title = {{Semiparametric estimators for the regression coefficients in the linear transformation competing risks model with missing cause of failure}},
	volume = {92},
	issn = {0006-3444},
	url = {https://doi.org/10.1093/biomet/92.4.875},
	doi = {10.1093/biomet/92.4.875},
	number = {4},
	urldate = {2025-11-04},
	journal = {Biometrika},
	author = {Gao, Guozhi and Tsiatis, Anastasios A.},
	month = dec,
	year = {2005},
	pages = {875--891},
}

@article{narasimhan_deconvolver_2020,
	title = {{{deconvolveR}: {a} {G}-{modeling} {program} for {deconvolution} and {empirical} {Bayes} {estimation}}},
	volume = {94},
	copyright = {Copyright (c) 2020 Balasubramanian Narasimhan, Bradley Efron},
	issn = {1548-7660},
	shorttitle = {{deconvolveR}},
	url = {https://doi.org/10.18637/jss.v094.i11},
	doi = {10.18637/jss.v094.i11},
	language = {en},
	urldate = {2025-11-06},
	journal = {Journal of Statistical Software},
	author = {Narasimhan, Balasubramanian and Efron, Bradley},
	month = sep,
	year = {2020},
	pages = {1--20},
}

@article{rubin_inference_1976,
	title = {{Inference and {missing} {data}}},
	volume = {63},
	issn = {0006-3444},
	url = {https://www.jstor.org/stable/2335739},
	doi = {10.2307/2335739},
	number = {3},
	urldate = {2025-11-11},
	journal = {Biometrika},
	author = {Rubin, Donald B.},
	year = {1976},
	note = {},
	pages = {581--592},
}

@article{juraska_quantifying_2026,
  title = {Quantifying how {HIV}-1 Envelope Sequence Features Impact Vaccine Efficacy in the {HVTN} 705/{HPX2008} Trial},
  author = {Juraska, Michal and Li, L. and Magaret, Craig A. and Giorgi, Elena E. and Molitor, C. and Ludwig, J. and Carpp, Lindsay N. and Peng, J. and DeCamp, Allan C. and Kee, Jia J. and Edlefsen, Paul T. and Rossenkhan, R. and Westfall, D. H. and Deng, W. and Chen, L. and Zhao, H. and Yssel, A. and Matten, D. and York, T. and Gwashu-Nyangiwe, A. and Ndabambi, N. and Thebus, R. and Swann, Ellen M. and Hural, John A. and van Duijn, Joost and Buchbinder, Susan and Tomaka, Frank and Corey, Lawrence and Mngadi, Khethokuhle and Gray, Glenda E. and Gilbert, Peter B. and Williamson, Carolyn and Mullins, James I.},
  journal = {eBioMedicine},
  year = {2026},
  note = {Accepted for publication}
}

\clearpage
\makeatletter
\global\backmatterfalse 
\@addtoreset{assumption}{section}
\@addtoreset{remark}{section}
\makeatother
\setcounter{section}{0}
\setcounter{equation}{0}
\setcounter{figure}{0}
\setcounter{table}{0}
\setcounter{assumption}{0}
\setcounter{remark}{0}
\renewcommand{\thesection}{Web Appendix \Alph{section}}
\renewcommand{\thesubsection}{\Alph{section}.\arabic{subsection}}
\renewcommand{\theequation}{S.\arabic{equation}}
\renewcommand{\thefigure}{S\arabic{figure}}
\renewcommand{\thetable}{S\arabic{table}}
\renewcommand{\theassumption}{\Alph{section}.\arabic{assumption}}
\renewcommand{\theremark}{\Alph{section}.\arabic{remark}}

\begin{center}
    {\LARGE\bfseries Supplementary Materials}
\end{center}
\vspace{8pt}
\noindent\textbf{Overview.} 
In Web Appendix A, we present two extensions to our method: allowing categorization into more than two bins and handling missing sequencing data. In Web Appendix B, we discuss alternatives to Assumption 4 (sequence depth conditional independence). In Web Appendix C, we discuss a strategy to select the threshold $q_0$. In Web Appendix D, we propose a pre-screening procedure for selecting binary features to test. In Web Appendix E, we provide additional intuition for how the classification model acts as a shrinkage estimator through a visual example. In Web Appendix F, we provide more details on our simulation study and include two additional simulations.
    
    The R code used for the simulations and data analysis is available at \url{https://github.com/jpspeng/multiseq_cox_sieve}.

\newpage 

\section{Methodology extensions}

\subsection{Categorization of $>2$ bins} \label{sec:appendix_bin}

Instead of having two bins defined by a single cutoff threshold $q_0$, suppose that we would like to define our failure type into $l$ bins defined by cut points $q :=  \{q_0, q_1, ..., q_l \}$, where $q_0 = 0$, $q_l = 1$, and $l > 2$. For individual $i$ who acquires HIV-1, we now define the mark variable $J_{i,q} \in \{1,...,l\}$ as an $l$-level categorical variable based on whether the true proportion $Q_i$ is within each pair of cut points:
\begin{equation}
    J_{i,q} :=
\begin{cases} 
1, & \text{if } q_0 \le Q_i \le q_1  \\ 
2, & \text{if } q_1< Q_i \le q_2 \\ 
... \\ 
l & \text{if } q_{l-1}< Q_i \le q_l
\end{cases}
\end{equation}

A potential example of interest would be to set cutpoints $\{q_1, q_2\}$ such that $q_1$ is close to 0 and $q_2$ is close to 1.

Our estimands of interest are now vaccine efficacy $VE_j$ against viral quasispecies with mark $J_q = j$ for $j \in \{1,2,...,l \}$. To facilitate estimation and inference with a Cox model in this setting, we now assume a proportional hazards assumption for the $l$ failure types: 
\begin{equation}
\begin{aligned}
    \lambda_{1s}(t; z, x) &= \exp\big(\theta_1^\top w\big)\,\lambda_{0,1s}(t), \\ 
    \lambda_{2s}(t; z, x) &= \exp\big(\theta_2^\top w\big)\,\lambda_{0,2s}(t), \\ 
    &\;\vdots \\ 
    \lambda_{ls}(t; z, x) &= \exp\big(\theta_l^\top w\big)\,\lambda_{0,ls}(t),
\end{aligned}
\end{equation}
where $\lambda_{0,js}(t)$ is the strata-specific baseline hazard for each failure type $j=1,\dots,l$ and $\theta_j$: $j \in \{1, ..., l \}$ is the vector of parameters to estimate. We use the same modified estimating equation, for $j \in \{1, ..., l \} $,  as Equation \eqref{eq:modified_estimating_eq}: 
\begin{align*}
U^{adj}_j(\theta_j) \;=\; \sum_{i=1}^n \int_0^\tau 
\Big\{ W_i - \overline{W}_{j,S_i}(t;\theta_j) \Big\} \, \nu_{q}(j; M_i, K_i, W_i, S_i, \widetilde{T}_i) \, dN_{i}(t),
\end{align*}
where the classification probabilities $\nu_{q}(j; M_i, K_i, W_i, S_i, \widetilde{T}_i)$ are now defined for each $j \in \{1, ..., l\}$. As with before, each event contributes to the estimating equation for each failure type $j \in \{1, ..., l\}$ weighted by the probability that the event was that failure type. To estimate this probability, we follow the same procedure as described in \ref{sec:classification_summary}, except that equation \eqref{eq:classification_prob} is replaced with
    \begin{equation}
        \widehat{\nu}_{q}(j; M_i, K_i, B_i) = \int_{q_{j-1}}^{q_j} \widehat{f}_{Q|M,K}(q|M_i, K_i, B_i, \Delta_i = 1) dq
    \end{equation}
We note that $\sum_{j=1}^l \nu_q(j;\,\cdot) = 1$.

\subsection{Missing sequencing data}
\label{sec:appendix_missing}

In some settings, sequencing data may be missing for a subset of
individuals who acquire HIV-1 during follow-up. We describe an approach
for estimation and inference in the presence of missing sequence
information using inverse probability weighting (IPW) and augmented
inverse probability weighting (AIPW)
\citep{gao_semiparametric_2005}. Let $R$ indicate whether an
individual's sequence information, $(K,M)$, is obtained, and let $A$
denote a vector of auxiliary covariates predictive of missingness. We
observe $(W_i,S_i,\widetilde{T}_i,A_i)$ for individuals with
$\Delta_i=0$ and $(W_i,S_i,\widetilde{T}_i,A_i,R_i,R_iK_i,R_iM_i)$
for individuals with $\Delta_i=1$.

For endpoint cases, let $O_i:=(W_i,S_i,A_i,\widetilde{T}_i)$
denote the variables observed regardless of whether sequencing is
obtained. We assume that sequencing information is missing at random
(MAR) \citep{rubin_inference_1976}, so that
\begin{equation}
P(R=1\mid J_{q_0},O,\Delta=1)
=
P(R=1\mid O,\Delta=1)
=:
\pi(O).
\label{eq:sequencing_mar}
\end{equation}
We also require the usual positivity condition: for some constant
$\epsilon>0$,
\begin{equation}
\pi(O)\geq\epsilon
\quad\text{almost surely among individuals with }\Delta=1.
\label{eq:sequencing_positivity}
\end{equation}

For sequenced endpoint cases, define $
\nu_{q_0}^{R}(j;M,K,O)
:=
P(J_{q_0}=j\mid M,K,O,R=1,\Delta=1).$
This probability can be estimated using the classification model
procedure described in Section~\ref{section:classification}. 

Define the observed score component
\begin{equation}
H_{ij}(\theta_j)
:=
\int_0^\tau
\left\{
W_i-\overline{W}_{j,S_i}(t;\theta_j)
\right\}
\,dN_i(t),
\label{eq:observed_score_component}
\end{equation}
where $N_i(t)$ is the counting process for an endpoint of either failure
type.

The IPW estimating function is
\begin{equation}
U_j^{\mathrm{IPW}}(\theta_j)
=
\sum_{i:\,\Delta_i=1,\,R_i=1}
\frac{H_{ij}(\theta_j)}{\pi(O_i)}
\nu_{q_0}^{R}(j;M_i,K_i,O_i).
\label{eq:ipw_est_eqn}
\end{equation}
Endpoint cases with missing sequencing information do not contribute
directly to this estimating equation. Instead, the inverse probability
weights re-balance the sequenced cases to represent the full population
of endpoint cases.

To construct an AIPW estimator, define the outcome regression $m_j(O_i)
:=
P(J_{i,q_0}=j\mid O_i,\Delta_i=1).$
Under the MAR assumption, $m_j(O) = \mathbb{E} \left\{
\nu_{q_0}^{R}(j;M,K,O)
\mid O,R=1,\Delta=1
\right\}$. Therefore, $m_j(O)$ can be estimated by regressing the estimated
classification probabilities
$\widehat{\nu}_{q_0}^{R}(j;M_i,K_i,O_i)$ on $O_i$ among sequenced
endpoint cases and predicting for all endpoint cases.

The AIPW estimating function is
\begin{align}
U_j^{\mathrm{AIPW}}(\theta_j)
&=
\sum_{i:\,\Delta_i=1}
H_{ij}(\theta_j)m_j(O_i)
\notag\\
&\quad+
\sum_{i:\,\Delta_i=1,\,R_i=1}
\frac{H_{ij}(\theta_j)}{\pi(O_i)}
\left\{
\nu_{q_0}^{R}(j;M_i,K_i,O_i)-m_j(O_i)
\right\}.
\label{eq:aipw_est_eqn}
\end{align}

Provided that the classification model is correctly specified, the
AIPW estimator is doubly robust with respect to the missingness and
outcome regression models. Under the stated MAR and positivity
assumptions, it remains consistent if either $\pi(O)$ or $m_j(O)$ is
correctly specified. Estimation and inference can be performed as follows:
\begin{enumerate}
\item Estimate the missingness probability
$\pi(O)=P(R=1\mid O,\Delta=1)$ among endpoint cases.

\item Estimate
$\nu_{q_0}^{R}(j;M,K,O)$ among sequenced endpoint cases using the
classification model procedure described in
Section~\ref{section:classification}.

\item Estimate $m_j(O)$ by regressing
$\widehat{\nu}_{q_0}^{R}(j;M,K,O)$ on $O$ among sequenced endpoint
cases and predicting for all endpoint cases.

\item Construct either
$U_j^{\mathrm{IPW}}(\theta_j)$ using
Equation~\eqref{eq:ipw_est_eqn} or
$U_j^{\mathrm{AIPW}}(\theta_j)$ using
Equation~\eqref{eq:aipw_est_eqn}, and solve the corresponding
estimating equation for $\widehat{\theta}_j$.

\item Estimate the variance using a nonparametric bootstrap that
re-estimates the classification, missingness, and outcome-regression
models within each bootstrap sample.
\end{enumerate}

Under the identification and modeling assumptions stated in the main
text and standard regularity conditions, the IPW estimator is
consistent when the classification and missingness models are
correctly specified. The AIPW estimator is consistent when the
classification model is correctly specified and either the
missingness model or the outcome regression model is correctly
specified. 

\newpage

\section{Alternatives to Assumption 4} 
\label{sec:appendix_assumption4}

When estimating the conditional density $f_Q(q \mid B,\Delta=1)$ across conditioning groups indexed by $B=(W,S,\widetilde{T})$, some groups may contain very few observations, which can lead to unstable or poorly identified estimates of the mixing distribution. To address this, we consider stronger versions of the conditional independence assumption that reduce the size of the conditioning set and simplify the estimation of $f_Q(q \mid B,\Delta=1)$. In this section, we present a general form of the assumption, examine several choices of the conditioning set, and provide sufficient conditions under which the corresponding conditional independence assumptions hold.

\begin{assumption}[General form of sequence depth conditional independence]
\label{assumption:sequence_general}
Let $B$ be a user-chosen subset of $(W,S,\widetilde{T})$ and let $\widetilde{B}$ denote the remaining components of $(W,S,\widetilde{T})$. Among observed failures,
\[
Q \perp (M,\widetilde{B}) \mid B,\Delta=1.
\]
\end{assumption}

Under Assumption~\ref{assumption:sequence_general}, the density needed in the classification model simplifies from $f_{Q \mid M}(q \mid M,W,S,\widetilde{T},\Delta=1)$ to $f_Q(q \mid B,\Delta=1)$, reducing the dimension of the conditioning set over which the mixing distribution of $Q$ must be estimated.

\subsection*{Case 1: $B=(W,S,\widetilde{T})$}

Setting $B=(W,S,\widetilde{T})$ yields the least restrictive version of the assumption:
\[
Q \perp M \mid W,S,\widetilde{T},\Delta=1,
\]
which is identical to Assumption~\ref{assumption:seq_depth_indep} in the main text. This assumption allows $Q$ to be associated with failure time and requires only that sequencing depth provide no additional information about $Q$ after conditioning on $(W,S,\widetilde{T})$.

In this case, the conditional distribution $f_Q(q\mid W,S,\widetilde{T},\Delta=1)$ must be modeled as a function of all components of $(W,S,\widetilde{T})$. If $\widetilde{T}$ is continuous, this requires an appropriate smoothing or discretization strategy rather than separate estimation at each observed failure time.

\subsection*{Case 2: $B=(W,S)$}

A more restrictive but potentially more practical choice is $B=(W,S)$, which corresponds to the assumption
\[
Q \perp (M,\widetilde{T}) \mid W,S,\Delta=1.
\]
Under this assumption, we only need to estimate $f_Q(q\mid W,S,\Delta=1)$, reducing the complexity of the deconvolution problem. The following conditions are sufficient for this independence assumption:

\begin{enumerate}
    \item[\textbf{(i)}] \textbf{Depth non-informativeness.}
    Assume that
    \[
    Q \perp M \mid W,S,\widetilde{T},\Delta=1.
    \]

    \item[\textbf{(ii)}] \textbf{Proportional $Q$-specific hazards.}
    Let $\lambda_s(t,q;w)$ denote the continuous mark-specific hazard density for a failure at time $t$ with mismatch proportion $Q=q$, conditional on $W=w$ and $S=s$. Assume that
    \[
    \lambda_s(t,q;w)=a_s(t;w)r_s(q;w),
    \]
    where $a_s(t;w)$ and $r_s(q;w)$ are nonnegative functions. Thus, failure hazards may differ across values of $Q$, but their relative magnitudes do not vary over time. Because the time-dependent factor $a_s(t;w)$ cancels when the hazard is normalized over $q$, the conditional distribution of $Q$ among failures does not depend on failure time. Under the censoring condition stated above, this implies
    \[
    Q \perp \widetilde{T}\mid W,S,\Delta=1.
    \]
\end{enumerate}

Together, conditions~(i)--(ii) imply
\[
Q \perp (M,\widetilde{T})\mid W,S,\Delta=1.
\]

\subsection*{Case 3: $B=Z$}

Because $W=(Z,X)$, choosing $B=Z$ corresponds to assuming
\[
Q \perp (M,\widetilde{T},X,S)\mid Z,\Delta=1.
\]
This is the most restrictive version of Assumption~\ref{assumption:sequence_general}, but it may be appropriate when $(X,S)$ primarily capture risk factors unrelated to the viral quasispecies composition. The following conditions are sufficient:

\begin{enumerate}
    \item[\textbf{(i)}] \textbf{Depth non-informativeness.}
    Assume that
    \[
    Q \perp M\mid Z,X,S,\widetilde{T},\Delta=1.
    \]

    \item[\textbf{(ii)}] \textbf{Proportional $Q$-specific hazards.}
    As in Case~2, assume that
    \[
    \lambda_s(t,q;z,x)=a_s(t;z,x)r_s(q;z,x).
    \]
    Under the censoring condition stated above, this implies
    \[
    Q \perp \widetilde{T}\mid Z,X,S,\Delta=1.
    \]

    \item[\textbf{(iii)}] \textbf{No residual association with adjustment covariates or stratum.}
    Assume that
    \[
    Q \perp (X,S)\mid Z,\Delta=1.
    \]
    Thus, among observed failures, $(X,S)$ provide no additional information about the mismatch proportion after conditioning on treatment assignment.
\end{enumerate}

Together, conditions~(ii)--(iii) imply
\[
Q \perp (\widetilde{T},X,S)\mid Z,\Delta=1.
\]
Combining this result with condition~(i) yields
\[
Q \perp (M,\widetilde{T},X,S)\mid Z,\Delta=1.
\]

The three cases reflect a trade-off between the strength of the assumptions and the complexity of estimating the mixing distribution. Case~1 uses the largest conditioning set and makes the weakest independence assumption, allowing the distribution of $Q$ to vary with failure time. Case~2 removes failure time from the conditioning set and is easier to implement, but requires the relative hazards associated with different values of $Q$ to remain stable over follow-up. Case~3 additionally removes $(X,S)$ and therefore requires these variables to provide no information about $Q$ beyond treatment assignment among observed failures. In practice, the largest conditioning set that permits stable estimation should be preferred, with smaller conditioning sets used only when their stronger assumptions are scientifically plausible.

\newpage

\section{Threshold $q_0$ selection}
\label{sec:threshold}

While the goal of our analysis may be to test for the presence of any mismatch, the resolution of our data is limited by the sequencing depth of samples. Therefore, we may wish to specify a lower threshold $q_0$ with higher sequencing depths and a higher $q_0$ with lower depths. Specifically, we may wish to set our threshold $q_0$ to be the smallest mismatch proportion detectable, akin to the limit of detection (LOD) in assays. For a given sequencing depth, we can define the LOD as the minimum true mismatch proportion such that there is at least some pre-specified (e.g. $\ge80\%$) probability of detection (POD) of the mismatch. This can be calculated as
$$
LOD = 1 - (1 - POD)^{1 / depth}.
$$
Table \ref{tab:lod_table} presents the LOD for various combinations of sequencing depth and POD.

\begin{table}[H]
\centering
\begin{tabular}{|c|c|c|c|}
\hline
\textbf{Sequence Depth} & \textbf{LOD (60\% POD)} & \textbf{LOD (80\% POD)} & \textbf{LOD (95\% POD)} \\
\hline
5    & 0.167 & 0.275 & 0.451 \\
10   & 0.088 & 0.149 & 0.259 \\
50   & 0.018 & 0.032 & 0.058 \\
100  & 0.009 & 0.016 & 0.030 \\
500  & 0.002 & 0.003 & 0.006 \\
1000 & 0.001 & 0.002 & 0.003 \\
\hline
\end{tabular}
\caption{Limits of detection (LOD) for different sequencing depths and probabilities of detection (POD).}
\label{tab:lod_table}
\end{table}

In our data, each sample has its own sequencing depth and, consequently, its own limit of detection. However, our method requires the specification of a single detection threshold $q_0$ for the entire dataset. Therefore, we define $q_0$ as the maximum of the treatment arm-specific LODs computed at the median sequencing depth within each treatment arm. For example, if the median sequencing depths are 50 in the vaccine arm and 100 in the placebo arm, the respective LODs (assuming an 80\% POD) are approximately 3.2\% and 1.6\%. In this case, we set $q_0 = 0.032$ so that the chosen threshold reflects the most conservative detection limit between study arms. For some binary features, we may also wish to consider another threshold $q_0$ set as 1 minus this threshold (e.g., in our running example, this would be the presence of a non-mismatch).  
\newpage

\section{Screening for viable marks}
\label{sec:screening_criteria}

In practice, sequence datasets may include hundreds or even thousands of binary features. Testing all features simultaneously in a single multi-sequence analysis would require a substantial multiplicity adjustment, which can severely diminish statistical power. 
To mitigate this issue, we propose a two-step screening procedure, agnostic to treatment assignment, to reduce the feature set prior to analysis. In the first step, we exclude marks exhibiting insufficient \emph{inter-individual} variability, as these provide little power to detect a sieve effect even under the most ideal conditions \citep{tarone_modified_1990}. 
In the second step, we remove marks with insufficient \emph{intra-individual} variability; such features can be adequately analyzed using standard single sequence sieve methods. 
Both screens depend on the chosen threshold $q_0$ defining the failure type  $J_{i,q_0}$.

One way to implement the first screen is as follows. We first restrict attention to cases with observed infections ($\Delta = 1$). Then, without examining treatment-arm–specific mark distributions, we assess whether a feature could in principle exhibit sufficient separation between treatment arms to have a detectable sieve effect under the most favorable configuration. Because the time-to-event sieve analysis involves considerably more structure—risk sets, censoring, covariate adjustment, and classification probabilities— using a simple cross-sectional comparison of case counts can ease calculation. To implement this, we identify the minimum number of primary endpoints of one virus type in one treatment arm (with zero of that type in the other arm) required for the corresponding Fisher’s exact test p-value to fall below a chosen significance threshold (e.g., 0.05). Although the true virus types $J_{i,q_0}$ are unknown and potentially misclassified, we can use the naive labels $\widetilde{J}_{i,q_0} = I(K_i/M_i \ge q_0)$ for the screening. We retain only those marks for which both 
$n_{\Delta = 1, \widetilde{J} = 1} = \sum_{i=1}^n I(\Delta_i = 1, \widetilde{J}_{i,q_0} = 1)$
and 
$n_{\Delta = 1, \widetilde{J} = 0} = \sum_{i=1}^n I(\Delta_i = 1, \widetilde{J}_{i,q_0} = 0)$
exceed this threshold, thereby eliminating features that lack sufficient power to detect a sieve effect.

\begin{figure}[h!]
    \centering
    \includegraphics[width=\textwidth]{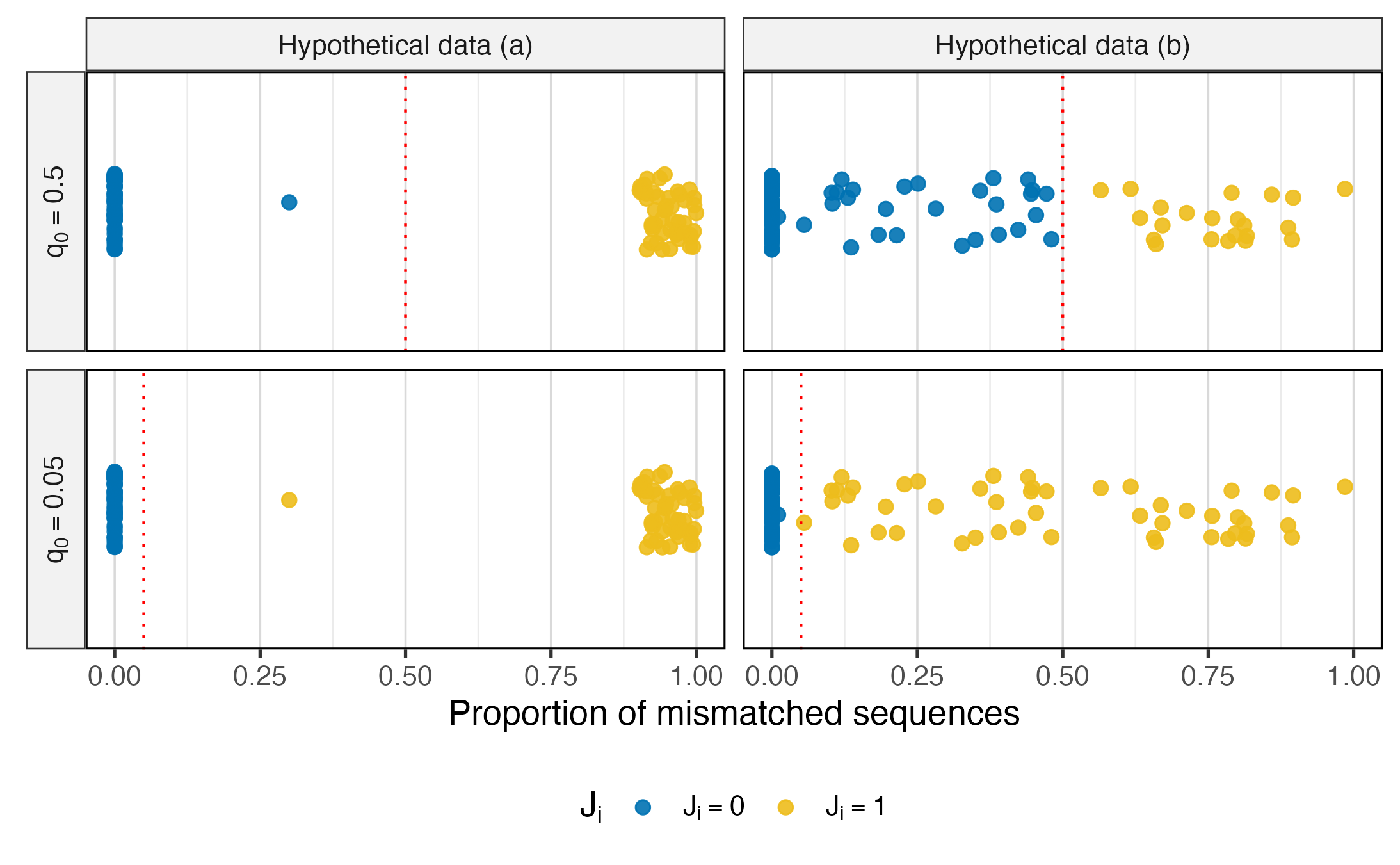}
    \caption{Illustration of the second screening step based on intra-individual diversity. 
Panels (a) and (b) show hypothetical datasets in which each point represents the proportion 
of mismatched sequences for an individual, with colors indicating the resulting 
virus-type classification. The dashed red line denotes the threshold $q_0$. In scenario (a), 
changing $q_0$ from 0.5 (top) to 0.05 (bottom) produces nearly identical classifications, 
indicating that a multi-sequence analysis cannot provide additional information beyond that provided from a modal mark ($q_0 = 0.5$) approach. In scenario (b), a substantial proportion of individuals change 
classification under the lower threshold, indicating that the multi-sequence method may 
meaningfully differ from the single-sequence analysis for this mark.}
    \label{fig:intra_variability_screen}
\end{figure}

In the second screen, we exclude marks for which there is insufficient intra-individual diversity relative to the $q_0$ threshold, such that the resulting multi-sequence analysis would be expected to yield conclusions similar to a single sequence sieve analysis. An illustration is given in Figure \ref{fig:intra_variability_screen}. In hypothetical data~(a), the virus classifications obtained from the modal sequence ($q_0 = 0.5$, top panel) and from applying a 5\% mismatch threshold ($q_0 = 0.05$, bottom panel) agree for all but one individual. In this setting, the multi-sequence analysis would not provide a meaningfully different result to the modal sequence analysis. In contrast, in hypothetical data~(b), a substantial proportion of individuals change failure-type classification when using the 5\% threshold, indicating that the multi-sequence analysis at that threshold may capture information not available from the modal sequence alone. As in the first screen, we rely on the naive labels $\widetilde{J}_{i,q_0}$ since the true virus types are unknown. A practical approach is to exclude marks for which fewer than a specified proportion (e.g., 10\%) of individuals are reclassified under the chosen threshold. 

\subsection{Screening for real data analysis}

In the data application presented in the manuscript, we applied two screening criteria. First, we screened for adequate inter-individual diversity: we required at least four primary endpoints with raw feature proportions $<1\%$ (or $<99\%$) and at least four with proportions $\ge 1\%$ (or $\ge 99\%$). Second, we required sufficient intra-individual diversity: the dichotomization induced by the 1\% (or 99\%) threshold had to differ from the dichotomization based on the mindist-sequence value in at least 10\% of primary endpoints.

\clearpage

\section{Classification model as a shrinkage estimator} \label{sec:appendix_shrinkage}

The estimator for the classification probabilities can be considered as a shrinkage estimator, combining the information from each observation with information from the entire sample. Equation \eqref{eq:classification_prob} can be written as follows:
    \begin{equation}
            \hat{f}_{Q \mid M, K}(q \mid M_i, K_i, B_i, \Delta_i = 1)
            = 
            \frac{
                f_{\text{binom}}(K_i; M_i, q)\,
               \widehat{f}_Q(q \mid B_i, \Delta_i = 1)
            }{
                \int_0^1 
                f_{\text{binom}}(K_i; M_i, q)\,
                 g(q ; \hat{\gamma}_{B_i})
                \, dq
            }.
        \label{eq:classiciation_probs_rewritten}
        \end{equation}
 In this equation, $f_{\text{binom}}(K_i; M_i, q)$ represents the information from the individual's observation, while $\widehat{f}_Q(q \mid B_i, \Delta_i = 1)$ represents the information from all the observations. If an observation's sequencing depth $M_i$ is large, the estimator $\widehat{f}_{Q|M,K}$ relies more on the observation rather than the sample. In contrast, if an observation's sequencing depth is small, the estimator $\widehat{f}_{Q|M,K}$ relies more on information from the entire sample's distribution \citep{whittemore_errors--variables_1989}.

\begin{figure}[h!]
    \centering
    \includegraphics[width=\textwidth]{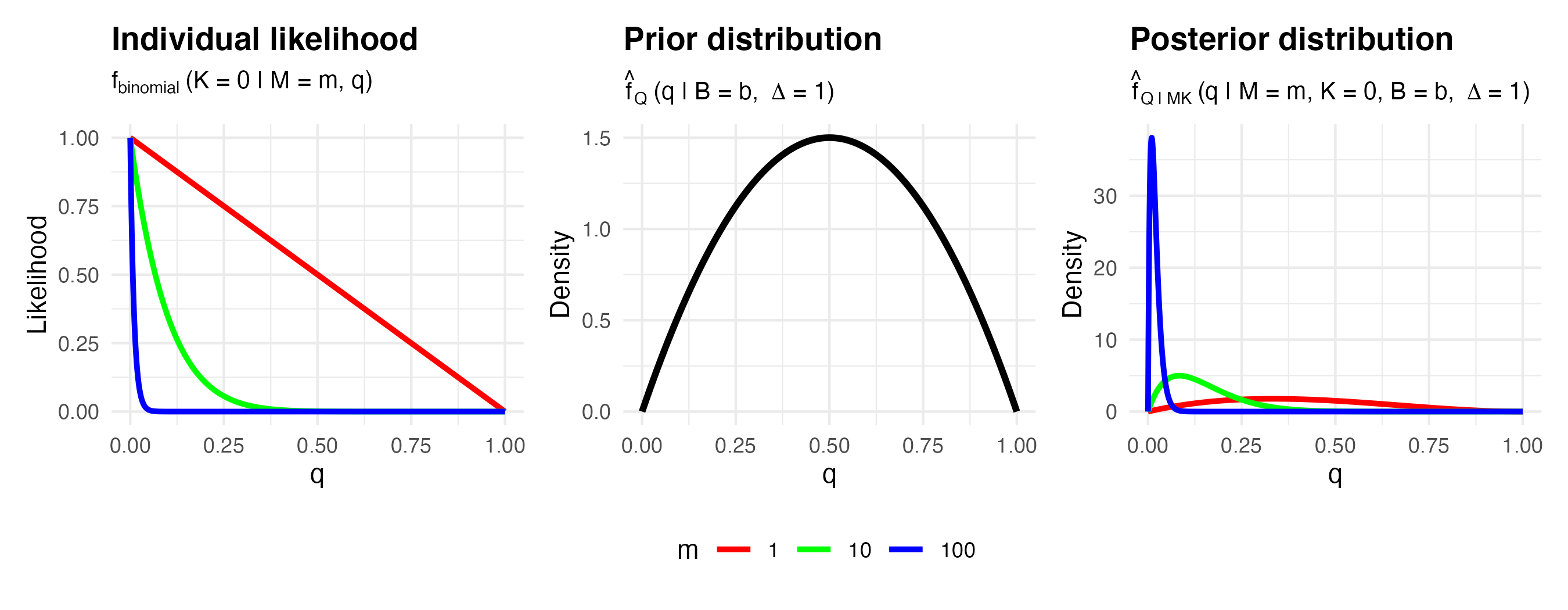}
    \caption{The components of equation \eqref{eq:classiciation_probs_rewritten} in a toy example.}
    \label{fig:shrinkage_demonstration}
\end{figure}

We can see this depicted in a simple example shown in Figure~\ref{fig:shrinkage_demonstration}. In this example, we restrict to a single stratum defined by $B = b$, and assume that the prior for $Q \mid B = b, \Delta = 1$ is estimated to be a $\text{Beta}(2, 2)$ distribution. Suppose we have three observations in this stratum, enumerated as $i = 1, 2, 3$, each with $K_i = 0$ mismatches but different sequencing depth: (1) $M_1 = 1$, (2) $M_2 = 10$, and (3) $M_3 = 100$. For each of these observations, we display the components of \eqref{eq:classification_prob}, including:  
\begin{itemize}
    \item $f_{\text{binom}}(K_i; M_i, q)$,
    \item the prior $\widehat{f}_Q(q \mid B = b, \Delta = 1)$, and
    \item the posterior $\widehat{f}_{Q|M,K}(q \mid M_i, K_i, B_i, \Delta_i = 1)$.
\end{itemize}

In the first observation with $M_1 = 1$ (red), the posterior $\widehat{f}_{Q|M,K}(q \mid M_i, K_i, B_i, \Delta_i = 1)$ (right panel) is only slightly changed from the prior $\widehat{f}_Q(q \mid B = b, \Delta = 1)$ (middle panel). However, for the third observation with sequencing depth $M_3 = 100$ (blue), the posterior is heavily weighted toward the observation (left panel) and differs greatly from the prior. This illustrates how the estimator combines individual- and sample-level information, with the relative influence of each determined by the sequencing depth $M_i$.

\clearpage

\section{Simulation Details and Additional Simulations} \label{sec:appendix_simulations}

\subsection{Details on Simulation Study Setup}

Failure times are generated with exponential distributions: $T_0 \sim \text{Exponential}(\gamma_0)$ and $T_1 \sim \text{Exponential}(\gamma_1)$, where the rate parameters depend on treatment assignment: $\gamma_0 \mid (Z, X) = 0.01 \exp(\beta_0 Z - 0.105 X)$ and $\gamma_1 \mid (Z, X) = 0.03 \exp(\beta_1 Z - 0.223 X)$. The two failure times are simulated independently, but there will be a   correlation between $T_0$ and $T_1$ for each individual induced by their shared vaccination status and covariate. If $T_0 < T_1$, the failure time is $T_0$ with failure type $J_{q_0} = 0$; otherwise, the failure time is $T_1$ with $J_{q_0} = 1$. We draw the true mismatch probabilities $Q$ using truncated Beta distributions as follows:
    $$
    Q \mid J_{q_0} \sim 
    \begin{cases}
    \text{Beta}(a = 0.5,\, b)\ \text{truncated to } [0,\, 0.01), & \text{if } J_{q_0} = 0, \\[6pt]
    \text{Beta}(a = 0.5,\, b)\ \text{truncated to } [0.01,\, 1], & \text{if } J_{q_0} = 1.
    \end{cases}
    $$
 The shape parameter $b$ of the Beta distribution differs for Settings (a)--(c) described below, so that the true marginal distribution of $Q|\Delta = 1$ arises approximately from a Beta distribution.  We assume administrative censoring at $t = 5$. With this setup, the probability of the event by $t=5$ in the placebo arm is roughly 15\%.

We consider three vaccine efficacy scenarios:
\begin{itemize}
    \item Setting (a) (no VE): $\beta_0 = \beta_1 = 0$, corresponding to 0\% VE for both viral types. For this setting, we set the Beta distribution's shape parameter $b = 5.7$.

    \item Setting (b) (positive VE, no sieve effect): $\beta_0 = \log(1 - 0.5)$ and $\beta_1 = \log(1 - 0.5)$, resulting in 50\% VE against both viral types. For this setting, we set the Beta distribution's shape parameter $b = 5.7$.

    \item Setting (c) (positive VE with sieve effect): $\beta_0 = \log(1 - 0.5)$ and $\beta_1 = \log(1 - 0.05)$, giving 50\% VE against viral type $J_{q_0} = 0$ and 5\% VE against viral type $J_{q_0} = 1$. For this setting, we set the Beta distribution's shape parameter $b = 3.8$.

\end{itemize}

\subsection{Simulation Study \#1 -- Results}

\begin{figure}[htbp!]
    \centering
    \includegraphics[width=\textwidth]{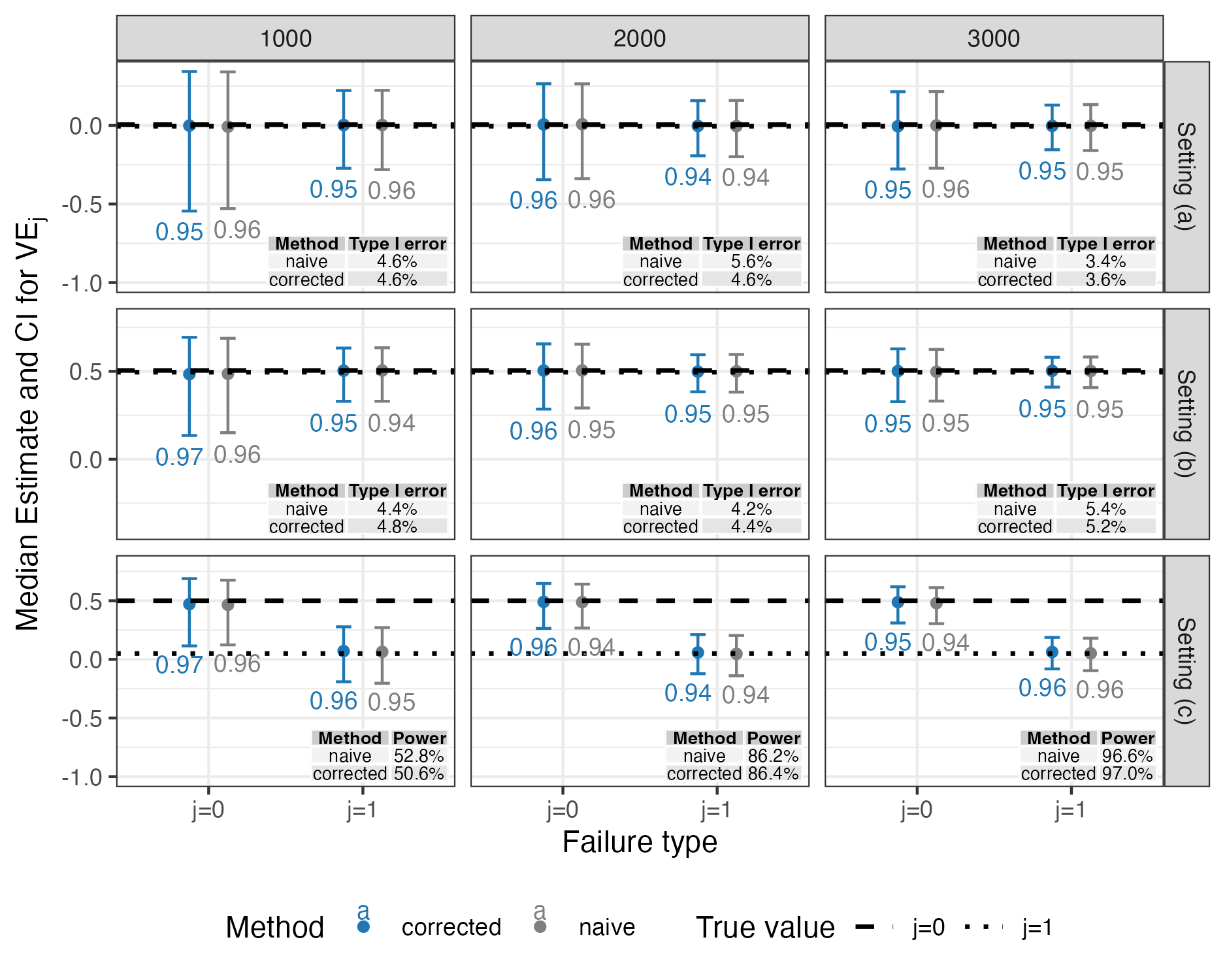}
    \caption{Results for Simulation Study \#1. The top, middle, and bottom panels display results for settings (a), (b), and (c) respectively. The left, middle, and right panels display results for differing sample sizes per treatment arm.  Each graph displays points for median $VE$ estimates for each failure type across all simulations, along with median lower and upper 95\% confidence interval bounds. The dashed and dotted horizontal lines are placed at the true values for the $j=0$ and $j=1$ failure types, respectively. The numbers below each error bar display the confidence interval coverage. We compare Type I error between the two estimators in settings (a) and (b) and power to detect a sieve effect in setting (c).}
    \label{fig:sim_fig1}
\end{figure}

Results for Simulation Study \#1, in which each endpoint case has a large sequencing depth ($M = 2{,}000$), are displayed in Figure \ref{fig:sim_fig1}. Across all three settings and sample sizes, the uncorrected estimator performs well: its median estimates are close to the true vaccine efficacies, and its nominal 95\% confidence intervals achieve near-nominal coverage. This is expected in this high sequencing depth scenario because misclassification of the empirical indicator $I(K_i/M_i \ge q_0)$ is extremely rare. The corrected estimator also performs well under these conditions, with point estimates and confidence intervals closely matching those of the uncorrected estimator. For settings (a) and (b), where the true vaccine efficacies are equal across viral types, both estimators exhibit appropriate type~I error rates when testing the null hypothesis $H_{B0}$. In setting (c), which includes a true sieve effect, both estimators show high power to reject $H_{B0}$, with power of 52.8\% (uncorrected) and 50.6\% (corrected) for 1{,}000 participants per treatment arm, 86.2\% (uncorrected) and 86.4\% (corrected) for 2{,}000 per arm, and 96.6\% (uncorrected) and 97.0\% (corrected) for 3{,}000 per arm.

\subsection{Simulation Study \#4}

In this additional simulation study, we confirm the properties of the estimator when the mixing distribution is specified correctly. Although this scenario is unrealistic for real data, it allows us to evaluate the estimator’s operating behavior under perfectly satisfied model assumptions.

\subsubsection{Data generating mechanism} 
Failure times are generated under a competing risks framework with two latent failure times. For each individual, the latent times for the two failure types are drawn independently as
$$
T_0 \sim \mathrm{Exponential}(\gamma_0), 
\qquad 
T_1 \sim \mathrm{Exponential}(\gamma_1),
$$
where the rate parameters depend on treatment assignment $Z$. Specifically,
$$
\gamma_0 = \kappa_Z (1 - p_Z), 
\qquad 
\gamma_1 = \kappa_Z p_Z,
$$
where $\kappa_Z$ controls the overall failure rate in treatment group $Z$, and $p_Z$ determines the relative frequency of the two failure types. In the simulations, we set $\kappa_0 = 0.04$ and $\kappa_1 = 0.03$, and we vary $p_z$ for $z \in \{0,1\}$ in Settings (a) and (b), as described below.

The failure type is defined as
$$
J_{q_0} =
\begin{cases}
0, & \text{if } T_0 < T_1, \\
1, & \text{if } T_1 < T_0 .
\end{cases}
$$
Administrative censoring is imposed at $t = 5$. 

For individuals who experience a failure ($\Delta = 1$), we generate the true mismatch probability $Q$ from truncated Beta distributions so that the failure type corresponds to whether $Q$ exceeds a fixed threshold $q_0 = 0.01$:
$$
Q \mid (J_{q_0}, Z) \sim
\begin{cases}
\mathrm{Beta}(\alpha_Z,\beta_Z) \text{ truncated to } [0, q_0], & J_{q_0} = 0, \\
\mathrm{Beta}(\alpha_Z,\beta_Z) \text{ truncated to } (q_0, 1], & J_{q_0} = 1.
\end{cases}
$$
The marginal distribution of $Q$ among failures in treatment group $Z = z$ is therefore $\mathrm{Beta}(\alpha_z,\beta_z)$.

We consider two configurations for the distribution of $Q$:
\begin{itemize}
\item \textbf{Setting (a):} $Q \mid (\Delta = 1, Z=z) \sim \mathrm{Beta}(1, 50)$ for both treatment groups. This corresponds to a true $VE_j = 0.25$ for both $j \in \{0, 1\}$. To generate this, we input $p_0 = p_1 = 1 - F_{\mathrm{Beta}(1,50)}(0.01) = 0.99^{50} \approx 0.605$.

\item \textbf{Setting (b):} $Q \mid (\Delta = 1, Z=0) \sim \mathrm{Beta}(1, 50)$ and $Q \mid (\Delta = 1, Z=1) \sim \mathrm{Beta}(1, 60)$. This corresponds to a true $VE_0 = 0.14$ and $VE_1 = 0.32$. To generate this, we input
$$
p_0 = 1 - F_{\mathrm{Beta}(1,50)}(0.01) = 0.99^{50} \approx 0.605,
\qquad
p_1 = 1 - F_{\mathrm{Beta}(1,60)}(0.01) = 0.99^{60} \approx 0.547.
$$
\end{itemize}

We consider the case in which sequencing depth variability differs between treatment arms. In the placebo arm, $M$ ranges uniformly from 1 to 15 in 20\% of endpoint cases and from 16 to 1{,}000 in the remaining cases. In the vaccine arm, $M$ ranges uniformly from 1 to 15 in 40\% of endpoint cases and from 16 to 1{,}000 in the remaining cases. Simulation sample sizes of $n = 1000$, $2000$, and $3000$ individuals per treatment arm are considered.

Across these sets of conditions, we compare bias, standard error, and confidence interval coverage between the proposed estimator and the uncorrected estimator (which uses the empirical, possibly incorrect indicator $\widetilde{J}_{i,q_0} = I(K_i / M_i \ge q_0)$ as the failure cause) over 1{,}000 simulations. We estimate the mixing distributions of $Q$ in each level of $Z$ with Beta distributions. Standard errors are estimated using 1{,}000 bootstrap samples, and Wald confidence intervals are constructed based on these estimates.

In Setting (a), the null hypothesis $VE_0 = VE_1$ holds. Therefore, we compare the type I error rates when testing the null $H_{B0}$ between the corrected and uncorrected methods. In Setting (b), the null hypothesis of equal vaccine efficacies across virus types does not hold, so we compare the power to detect the sieve effect and reject $H_{B0}$ between the two estimators.

\subsubsection{Simulation results}
\begin{figure}[htbp!]
    \centering
    \includegraphics[width=\textwidth]{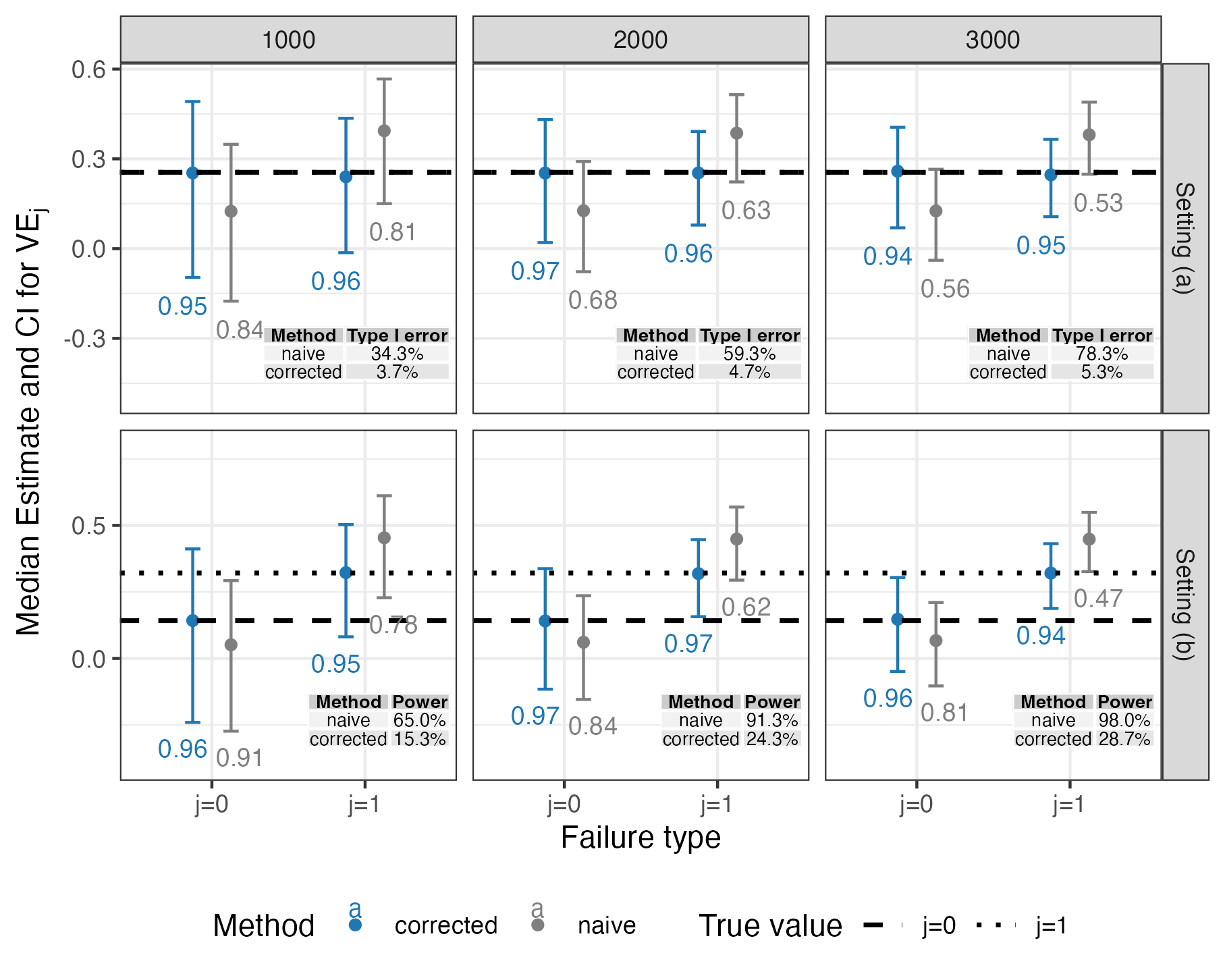}
    \caption{Results for Simulation Study \#4. The top and bottom panels display results for settings (a) and (b) respectively. The left, middle, and right panels display results for differing sample sizes per treatment arm. Each graph displays points for median $VE$ estimates for each failure type across all simulations, along with median lower and upper 95\% confidence interval bounds. The dashed and dotted horizontal lines are placed at the true values for the $j=0$ and $j=1$ failure types, respectively. The numbers below each error bar display the confidence interval coverage. We compare Type I error between the two estimators in (a) and power to detect a sieve effect in (b).}
    \label{fig:sim_fig4}
\end{figure}

Results are shown in Figure~\ref{fig:sim_fig4}. In Setting (a), where the null hypothesis of equal vaccine efficacies holds, the uncorrected estimator exhibits substantial bias in opposite directions for the two failure types: $VE_0$ is underestimated while $VE_1$ is overestimated. This produces confidence interval coverage below the nominal level, with coverage worsening as sample size increases. The resulting type I error rate is highly inflated, increasing from 34.3\% with $n=1000$ individuals per treatment arm to 78.3\% with $n=3000$ individuals per treatment arm. In contrast, the corrected estimator is approximately unbiased, achieves near-nominal confidence interval coverage, and maintains type I error close to the nominal 5\% level across sample sizes.

In Setting (b), where the null hypothesis of equal vaccine efficacies does not hold, the uncorrected estimator again shows substantial bias, underestimating $VE_0$ and overestimating $VE_1$. Although this leads to high apparent power to reject $H_{B0}$, the rejection is driven in part by bias in the estimated vaccine efficacies rather than accurate recovery of the true sieve effect. The corrected estimator remains closer to the true values and achieves near-nominal confidence interval coverage for both failure types. Its power to detect the sieve effect increases with sample size, from 15.3\% at $n=1000$ individuals per treatment arm to 28.7\% at $n=3000$ individuals per arm.

\subsection{Simulation Study \#5}

In the main simulation studies, we fixed the threshold at $q_0=0.01$. In this additional simulation study, we examine the sensitivity of the proposed estimator to the choice of $q_0$ by evaluating performance at $q_0 \in \{0.01,0.02,0.05\}$. This simulation uses a data-generating mechanism in which, under the sieve setting, vaccine efficacy varies with the underlying mismatch proportion $Q$, so that viral populations with lower and higher mismatch proportion have different failure rates.

\subsubsection{Data generating mechanism}

We use the same basic setup as in the main simulation studies, with two equal-sized treatment arms (2,000 participants each), placebo $(Z=0)$ and vaccine $(Z=1)$, and a binary covariate $X \sim \mathrm{Bernoulli}(0.5)$. Administrative censoring is imposed at $t=5$.

Instead of generating failure times from two discrete failure types, we generate failure times from a continuous model. Let $f_0(q)$ denote the distribution of $Q$ among placebo-arm failures, and let $h(q)$ denote the vaccine-arm hazard ratio for an infection with mismatch proportion $q$. The marginal vaccine-arm hazard ratio is
$$
A_1 = \int_0^1 h(q) f_0(q)\,dq,
$$
with $A_0=1$ for the placebo arm. Failure times are generated as
$$
T \sim \mathrm{Exponential}\left\{
\lambda_0 \exp(\alpha X) A_Z
\right\},
$$
where $\lambda_0=-\log(0.85)/5$, $\alpha=\log(0.9)$, and $A_Z=A_0$ for $Z=0$ and $A_Z=A_1$ for $Z=1$.

For individuals who experience a failure $(\Delta=1)$, $Q$ is drawn from $f_0(q)$ in the placebo arm and from
$$
f_1(q) =
\frac{
h(q)f_0(q)
}{
\int_0^1 h(u)f_0(u)\,du
}
$$
in the vaccine arm. Therefore, when $h(q)$ is larger for larger values of $q$, vaccine-arm failures are enriched for viral populations with higher mismatch proportions.

We specify the placebo failure-mark distribution as
$$
f_0(q)
=
0.45 \, \mathrm{Beta}(0.35,80)
+
0.35 \, \mathrm{Beta}(2,35)
+
0.20 \, \mathrm{Beta}(4,30).
$$
This distribution places substantial mass near zero while retaining non-negligible mass above each of the thresholds $0.01$, $0.02$, and $0.05$.

We consider two vaccine efficacy settings:
\begin{itemize}
\item \textbf{Setting (a): positive VE, no sieve effect.} We set $h(q)=0.58$ for all $q$, corresponding to a constant vaccine efficacy of 42\% for all values of the mismatch proportion.

\item \textbf{Setting (b): positive VE with a sieve effect.} We set
$$
h(q)
=
0.50 + 0.45 \, \mathrm{expit}\left\{
\frac{q-0.02}{0.01}
\right\}.
$$
This corresponds to stronger vaccine protection against viral populations with low mismatch burden and weaker vaccine protection against viral populations with higher mismatch burden.
\end{itemize}

For each simulated dataset, we evaluate the estimators separately at $q_0=0.01$, $0.02$, and $0.05$. The true vaccine efficacy for failure type $j$ at threshold $q_0$ is
$$
VE_j(q_0)
=
1 -
\frac{
\int_{\mathcal Q_j(q_0)} h(q) f_0(q)\,dq
}{
\int_{\mathcal Q_j(q_0)} f_0(q)\,dq
},
$$
where $\mathcal Q_0(q_0)=\{q:q<q_0\}$ and $\mathcal Q_1(q_0)=\{q:q\ge q_0\}$. In Setting (a), $VE_0(q_0)=VE_1(q_0)$ for all values of $q_0$. In Setting (b), the two vaccine efficacies differ, with the magnitude of the sieve effect depending on the threshold.

We use the same sequencing-depth mechanism as Simulation Study \#3, in which sequencing depth variability differs between treatment arms. In the placebo arm, $M$ ranges uniformly from 1 to 15 in 20\% of endpoint cases and from 16 to 1{,}000 in the remaining cases. In the vaccine arm, $M$ ranges uniformly from 1 to 15 in 40\% of endpoint cases and from 16 to 1{,}000 in the remaining cases. Given $M$ and $Q$, we generate $K \mid M,Q \sim \mathrm{Binomial}(M,Q)$.

Across these conditions, we compare bias, standard error, and confidence interval coverage between the proposed estimator and the uncorrected estimator at each value of $q_0$. For the proposed estimator, classification probabilities are estimated separately for each threshold using penalized spline mixtures with $\mathrm{df}=5$ and $c_0=0.01$, augmented with point mass basis functions at 0 and 1. We use the same number of simulation replicates, bootstrap procedure, and Wald confidence interval construction as in the main simulation studies.

In Setting (a), the null hypothesis $VE_0(q_0)=VE_1(q_0)$ holds for all values of $q_0$. Therefore, we compare type I error rates when testing $H_{B0}$ between the corrected and uncorrected methods. In Setting (b), the null hypothesis of equal vaccine efficacies does not hold, so we compare power to detect the sieve effect and reject $H_{B0}$ across the three threshold choices.

\subsubsection{Simulation results}

\begin{figure}[htbp!]
    \centering
    \includegraphics[width=\textwidth]{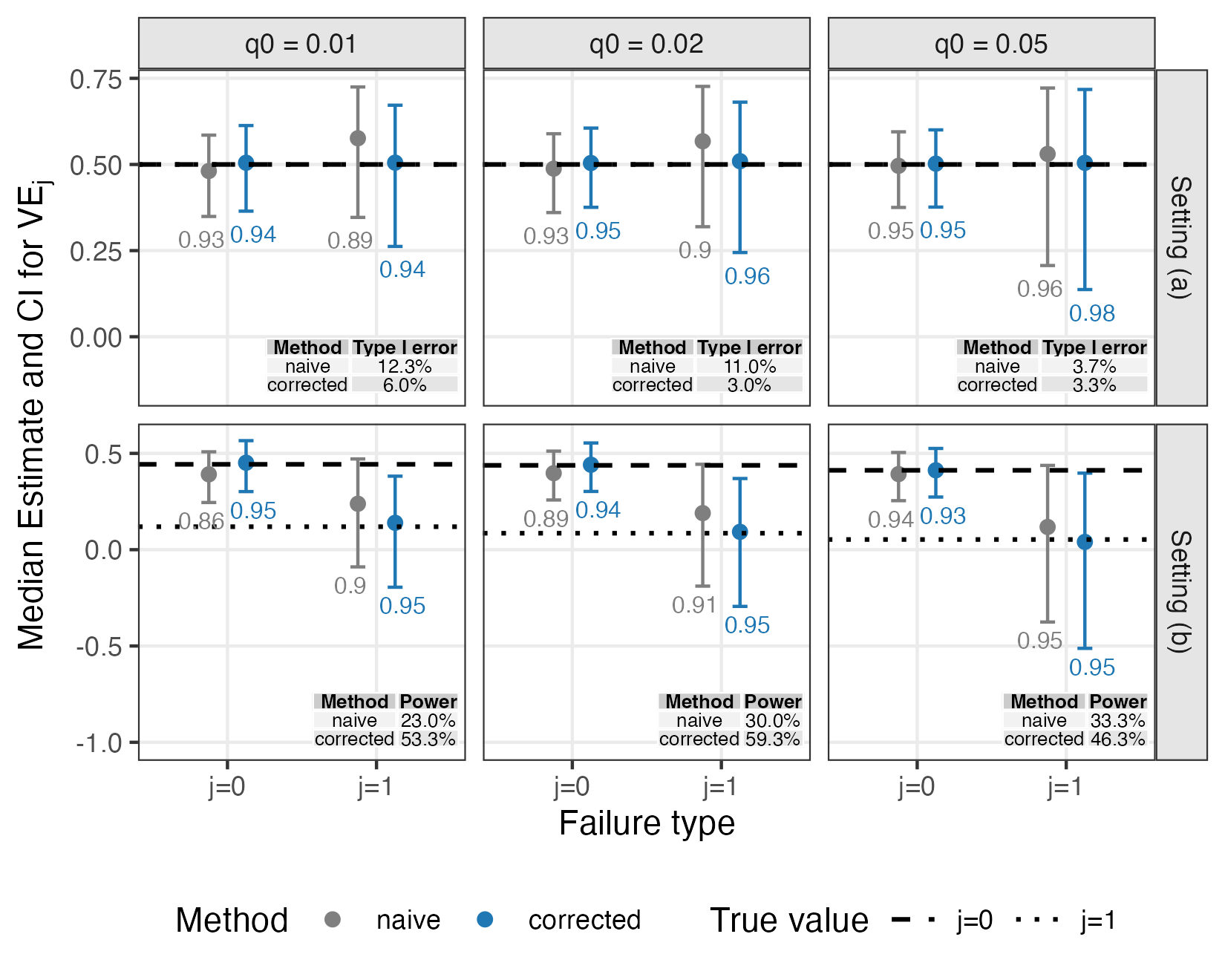}
    \caption{Results for Simulation Study \#5. The top and bottom panels display results for settings (a) and (b) respectively. The left, middle, and right panels display results for differing thresholds $q_0$. Each graph displays points for median $VE$ estimates for each failure type across all simulations, along with median lower and upper 95\% confidence interval bounds. The dashed and dotted horizontal lines are placed at the true values for the $j=0$ and $j=1$ failure types, respectively. The numbers below each error bar display the confidence interval coverage. We compare Type I error between the two estimators in (a) and power to detect a sieve effect in (b).}
    \label{fig:sim_fig5}
\end{figure}

Results are shown in Figure~\ref{fig:sim_fig5}. In Setting (a), where there is positive vaccine efficacy but no sieve effect, both estimators produce median estimates close to the true values across the three thresholds. However, the uncorrected estimator shows some inflation of the type I error rate at smaller thresholds, with type I error of 12.3\% and 11.0\% for $q_0=0.01$ and $q_0=0.02$, respectively. In contrast, the corrected estimator maintains type I error closer to the nominal 5\% level across thresholds.

In Setting (b), where vaccine efficacy varies with the underlying mismatch proportion $Q$, the uncorrected estimator attenuates the contrast between $VE_0(q_0)$ and $VE_1(q_0)$. This is most evident at smaller thresholds, where the uncorrected estimator tends to underestimate $VE_0(q_0)$ and overestimate $VE_1(q_0)$. The corrected estimator is closer to the true values and achieves near-nominal confidence interval coverage across most settings. The corrected estimator also has higher power to detect the sieve effect than the uncorrected estimator at all thresholds, with power of 53.3\%, 59.3\%, and 46.3\% for $q_0=0.01$, $0.02$, and $0.05$, respectively, compared to 23.0\%, 30.0\%, and 33.3\% for the uncorrected estimator. The larger confidence intervals for $VE_1(q_0)$ at higher thresholds reflect the smaller number of endpoint cases with $Q \ge q_0$.

\clearpage

\label{lastpage}
\end{document}